\documentclass[journal,onecolumn, draftcls, 12pt]{IEEEtran} 

\usepackage[utf8]{inputenc}
\usepackage[T1]{fontenc}
\usepackage[hidelinks]{hyperref}
\usepackage{url}
\usepackage{ifthen}
\usepackage{cite}
\usepackage[cmex10]{amsmath} 
\usepackage{amsfonts,amssymb}
\usepackage{graphicx}
\usepackage{color}
\usepackage{ifthen}
\usepackage{comment}
\usepackage[arxiv]{optional} 

\usepackage[ruled]{algorithm2e}
\usepackage{float}

\allowdisplaybreaks[1]

\usepackage{amsthm}

\begin{document}
\title{Mechanism Design for Demand Management in Energy Communities}

 \author{%
   \IEEEauthorblockN{Xupeng Wei, Achilleas Anastasopoulos\\}
   \IEEEauthorblockA{University of Michigan\\
                     Ann Arbor, MI 48109, USA\\
                     Email: \texttt{\{xupwei,anastas\}@umich.edu}}
 }

\maketitle




\optv{2col}{
}

\def\cA{\mathcal{A}}
\def\cB{\mathcal{B}}
\def\cC{\mathcal{C}}
\def\cE{\mathcal{E}}
\def\cF{\mathcal{F}}
\def\cG{\mathcal{G}}
\def\cN{\mathcal{N}}
\def\cP{\mathcal{P}}
\def\cR{\mathcal{R}}
\def\cS{\mathcal{S}}
\def\cT{\mathcal{T}}
\def\cU{\mathcal{U}}
\def\cV{\mathcal{V}}
\def\cW{\mathcal{W}}
\def\cX{\mathcal{X}}
\def\cY{\mathcal{Y}}
\def\cZ{\mathcal{Z}}
\def\cPi{\mathcal{\Pi}}

\def\tw{\tilde{w}}
\def\pih{\hat{\pi}}
\def\Pih{\hat{\Pi}}

\newtheorem{lemma}{Lemma}
\newtheorem{fact}{Fact}
\newtheorem{theorem}{Theorem}
\newtheorem{corollary}{Corollary}
\newtheorem{assumption}{Assumption}

\newcommand{\ve}[1]{\boldsymbol{#1}}
\newcommand{\eqdef}{\stackrel{\scriptscriptstyle \triangle}{=}}
\newcommand{\mdef}{\stackrel{\text{\tiny def}}{=}}
\newcommand{\E}{\mathbb{E}}
\def\Real{\mathbb{R}}
\def\P{\mathbb{P}}

\def\Tr{\mathsf{T}}
\def\Ra{\mathcal{RA}}
\def\cM{\mathcal{M}}
\def\cL{\mathcal{L}}
\def\Gr{\mathcal{GR}}
\def\one{\boldsymbol{1}}
\def\Rp{\mathcal{RP}}

\newcommand{\red}[1]{\textcolor{red}{#1}}

\begin{abstract}
       We consider a demand management problem of an energy community, in which several users obtain energy from an external organization such as an energy company, and pay for the energy according to pre-specified prices that consist of a time-dependent price per unit of energy, as well as a separate price for peak demand.
       Since users' utilities are their private information, which they may not be willing to share, a mediator, known as the planner, is introduced to help optimize the overall satisfaction of the community (total utility minus total payments) by mechanism design.
       A mechanism consists of a message space, a tax/subsidy and an allocation function for each user. Each user reports a message chosen from her own message space, and then receives some amount of energy determined by the allocation function and pays the tax specified by the tax function. A desirable mechanism induces a game, the Nash equilibria (NE) of which result in an allocation that coincides with the optimal allocation for the community.

       As a starting point, we design a mechanism for the energy community with desirable properties such as full implementation, strong budget balance and individual rationality for both users and the planner. We then modify this baseline mechanism for communities where message exchanges are allowed only within neighborhoods, and consequently, the tax/subsidy and allocation functions of each user are only determined by the messages from her neighbors. All the desirable properties of the baseline mechanism are preserved in the distributed mechanism. Finally, we present a learning algorithm for the baseline mechanism, based on projected gradient descent, that is guaranteed to converge to the NE of the induced game.
\end{abstract}

\newpage

\section{Introduction}

Resource allocation is an essential task in networked systems such as communication networks, energy/power networks, etc~\cite{schmidt2009distributed,nair2018multi,hamaali2021resources}. In such systems, there is usually one or multiple kinds of limited and divisible resources allocated among several agents. When full information regarding agents’ interests is available, solving the optimal resource allocation problem reduces to a standard optimization problem. However, in many interesting scenarios, strategic agents may choose to conceal or misreport their interests in order to get more resources. In such cases, it is possible that appropriate incentives are designed so that selfish agents are incentivized to report truly their private information, thus enabling optimal resource allocation~\cite{menache2011network}.

In existing works related to resource allocation problems, \emph{mechanism design}~\cite{hurwicz2006designing,borgers2015introduction} is frequently used to address the agents' strategic behavior mentioned above. In the framework of mechanism design, the participants reach an agreement regarding how they exchange messages, how they share the resources, and how much they pay (or get paid). Such agreements are designed to incentivize the agents to provide the information needed to solve the optimization problem.

In this paper, we develop mechanisms to solve a demand management problem in energy communities. In an energy community, users obtain energy from an energy company and pay for it. The pre-specified prices dictated by the energy company consist of a time-dependent price per unit of energy, as well as a separate price for peak demand. 
Users' demand is subject to constraints relating to equipment capacity and minimum comfort level. Each user possesses a utility as a function of their own demand. Utilities are private information for users. The welfare of the community is the sum of utilities minus energy cost. If users were willing to report truthfully their utilities, one could easily optimize energy allocation to maximize social welfare. However, since users are strategic and might not be willing to report utilities directly, to maximize the welfare, we need to find an appropriate mechanism that incentivizes them to reveal some information about their utilities, so that optimal allocation is reached even in the presence of strategic behaviors. These mechanisms are usually required to possess several interesting properties, among which, full implementation in Nash equilibria (NE), individual rationality and budget balance~\cite{borgers2015introduction,garg2008foundations,garg2008foundations2}. Moreover, in environments with communication constraints, it is desirable to have ``distributed'' mechanisms, whereby energy allocation and tax/subsidies for each user can be evaluated using only local messages in the user's neighborhood. 
Finally, for actual deployment of practical mechanisms we hope that the designed mechanism has convergence properties that guarantee that NE is reached by the agents by means of a provably convergent learning algorithm.

\subsection{Contributions}

This paper proposes a way of designing a mechanism for implementing the optimal allocation of the demand management problem in an energy community, where there are strategic users communicating over a pre-specified message exchange network. The main contributions of our work are as follows:

\begin{itemize}

\item We design a baseline, ``centralized'' mechanism for an environment with concave utilities and convex constraints.
A ``centralized'' mechanism allows messages from all users to be communicated to the planner~\cite{borgers2015introduction,garg2008foundations,garg2008foundations2}. To avoid excessive communication cost brought by direct mechanisms (due to messages being entire utility functions), the mechanisms proposed in this paper are indirect, non-VCG type~\cite{vickrey1961counterspeculation,clarke1971multipart,groves1973incentives}, with messages being real vectors with finite (and small) dimensionality.
Unlike related previous works~\cite{kelly1998rate,yang2005revenue,basar2006,SiAn17b}, a simple form of allocation function is adopted, namely, allocation equals demand.
The mechanism possesses the properties of full implementation, budget balance and individual rationality~\cite{borgers2015introduction,garg2008foundations,garg2008foundations2}.
Although we develop the mechanism for demand management in energy communities, the underlying ideas can be easily adapted to other problems and more general environments. Specifically, environments with non-monotonic utilities, external fixed unit prices, and the requirement of peak shaving are tractable with the proposed mechanism.

\item Inspired by the vast literature on distributed non-strategic optimization~\cite{rabbat2004distributed,boyd2011distributed,wei2013distributed,alvarado2014new,di2016distributed}, as well as our recent work on distributed mechanism design (DMD)~\cite{SiAn19,HeAn20}, we modify the baseline mechanism and design a ``distributed'' version of it. A distributed mechanism can be deployed in environments with communication constraints, where users' messages cannot be communicated to the central planner; consequently the allocation and tax/subsidy functions for each user should only depend on messages from direct neighbors. The focus of our methodology is to show how a centralized mechanism can be modified into a decentralized one in a systematic way by means of introducing extra message components that are acting as proxies of the messages not available to a user due to communication constraints. An added benefit of this systematic design is that the new mechanism preserves all the desirable properties of the centralized mechanism.

\item Since mechanism design (centralized or distributed) deals with equilibrium properties, one relevant question is how equilibrium is reached when agents enter the mechanism. Our final contribution in this paper is to provide a ``learning'' algorithm~\cite{brown1951iterative,monderer1996fictitious,hofbauer2002global,milgrom1990rationalizability,scutari2012monotone} that addresses this question for the case of the proposed centralized mechanism. The algorithm is based on the projected gradient descent method in optimization theory~\cite[Ch.~7]{polyak1987opt}. Learning proceeds through price adjustments and demand announcements according to the prices. During this process, users don't need to reveal the entire utility functions. Convergence of the message profile toward one NE is conclusively proved and since the mechanism is designed to fully implement the optimal allocation in NE, this implies that the allocation corresponding to the limiting message profile is the social welfare maximizing solution.
\end{itemize}

\subsection{Related Literature}

The model for demand management in energy communities investigated in this paper originates from network utility maximization (NUM) problems, which is one typical category of resource allocation problems in networks (see~\cite[Chapter~2]{srikant2013communication} for a detailed approach to models and algorithms for solving NUM problems).
There are two distinct research directions that have emanated from the standard centralized formulations of optimization problems.

The first direction addresses the problem of communication constraints when solving an optimization problem in a centralized fashion. Taking into account these communication constraints several researchers have proposed \emph{distributed optimization} methods~\cite{rabbat2004distributed,boyd2011distributed,wei2013distributed,alvarado2014new,di2016distributed} whereby an optimization problem is solved by means of message-passing algorithms between neighbors in a communication network. The works have been further refined to account for possible users' privacy concerns during the optimization processes~\cite{huang2015differentially,cortes2016differential,nozari2016differentially,han2017differentially}.
Nevertheless, the users are assumed to be non-strategic in this line of works.

The second research direction, namely \emph{mechanism design}, addresses the presence of strategic agents in optimization problems in a direct way.
The past several decades have witnessed applications of this approach in various areas of interest, such as market allocations~\cite{groves1977optimal,hurwicz1979outcome,yang2005revenue}, spectrum sharing~\cite{huang2006auction,wang2008game,wang2010toda}, data security~\cite{ghosh2011selling,khalili2017designing,pal2020data}, smart grid~\cite{caron2010incentive,samadi2012advanced,muthirayan2019mechanism}, etc.
The well-known VCG mechanism~\cite{vickrey1961counterspeculation,clarke1971multipart,groves1973incentives} has been utilized extensively in this line of research. In VCG, users have to communicate utilities (i.e., entire functions), which leads to a high cost of information transmission. To ease the burden of communication, Kelly's mechanism~\cite{kelly1998rate} (and extensions to multiple divisible resources~\cite{iosifidis2013iterative}) has been proposed as a solution, which uses logarithmic functions as surrogates of utilities. The users need only report a real number and thus the communication cost reduces dramatically, at the expense of efficiency loss~\cite{johari2004efficiency} and/or the assumption of price-taking agents.
A number of works extend Kelly's idea to reduce message dimensionality in strategic settings~\cite{yang2007vcg,johari2009efficiency,farhadi2018surrogate}.
Other indirect mechanisms guaranteeing full implementation in environments with allocative constraints have been proposed in~\cite{demos,kakhbodcorrection} using penalty functions to incentivize feasibility, and in~\cite{kelly1998rate,yang2005revenue,basar2006} using  \emph{proportional allocation}, or its generalization, \emph{radial allocation}~\cite{SiAn14b,SiAn17b}.
All aforementioned works on mechanism design can be categorized as ``centralized'' mechanisms, which means that
agents' messages are broadcasted to a central planner who evaluates allocation and taxation for all users.
The first attempts in designing decentralized mechanisms were reported in~\cite{SiAn19,HeAn20}, where mechanisms are designed with the additional property that the allocation and tax functions for each agent depend only on the messages emitted by neighboring agents. As such, allocation and taxation can be evaluated locally.

Finally, learning in games is motivated by the fact that NE is, theoretically, a complete information solution concept.
However, since users do not know each-others' utilities, they cannot evaluate the designed NE off line.
Instead there is a need for a process (learning) during which the NE is being learnt by the community.
The classic works~\cite{brown1951iterative,monderer1996fictitious,hofbauer2002global} adopt \emph{fictitious play}, while in~\cite{milgrom1990rationalizability} a connection between supermodularity and convergence of learning dynamics within an \emph{adaptive dynamics} class is made, and is further specialized in~\cite{chen2002family} to the Lindahl allocation problem.
A general class of learning dynamics named \emph{adaptive best response} is discussed in~\cite{healy2012designing} in connection with \emph{contractive games}.
Learning in \emph{monotone games}~\cite{scutari2012monotone,scutari2014real} is investigated in~\cite{gharesifard2013distributed,ye2016game,grammatico2017dynamic,yi2018distributed,paccagnan2018nash,parise2019variational}, with further applications in network optimization~\cite{xiao2019distributed,bimpikis2019cournot}.
Recently, learning of NE utilizing \emph{reinforcement learning} has been reported in~\cite{zhang2019policy,uz2020reinforcement,roudneshin2020reinforcement,shi2020multi,sohet2020learning}.


\section{Model and Preliminaries}

\subsection{Demand Management in Energy Communities}

Consider an energy community consisting of $N$ users and a given time horizon $T$, where $T$ can be viewed as the number of days during one billing period.
Each user $i$ in the user set $\cN$ has her own prediction on her usage over one billing period denoted by $\ve{x}^i = (x^i_1,\ldots,x^i_T)$, where $x^i_t$ is the predicted usage of user $i$ on the $t$-th time slot of the billing period\footnote{Throughout the paper we use superscripts to denote users and constraints and subscripts to denote time slots.}.
Note that $x^i_t$ can be a negative number due to the potential possibility that users in the electrical grid can generate power through renewable technologies (e.g., photovoltaic) and return the surplus back to the grid.
The users are characterized by their utility functions as
\begin{equation*}
       v^i(\ve{x}^i) = \sum_{t=1}^{T} v^i_t(x^i_t), \forall i \in \cN.
\end{equation*}

The energy community, as a whole, pays for the energy.
The unit prices are given separately for every time slot $t$ denoted by $p_t$.
These prices are considered given and fixed (e.g., by the local utility company).
In addition, the local utility company imposes a unit peak price $p_0$ in order to incentivize load balancing and lessen the burden of peaks in demand.
To conclude, the cost of energy to the community is as follows:
\begin{equation}
       \label{eq:cost}
       J(\ve{x}) = \sum_{t=1}^T p_t \left( \sum_{i=1}^N x^i_t \right)
        + p_0 \cdot \underset{1 \leq t \leq T}{\max} \sum_{i=1}^N x^i_t,
\end{equation}
where $\ve{x}$ is a concatenation of demand vectors $\ve{x}^1,\ldots,\ve{x}^N$.

The centralized demand management problem for the energy community can be formulated as
\begin{equation}
\label{opt:woassump}
    \underset{\ve{x} \in \cX}{\text{maximize}}\quad \sum_{i=1}^N v^i(\ve{x}^i)-J(\ve{x}).
\end{equation}
The meaning of the feasible set $\cX$ is to incorporate possible lower bounds on each user's demand (e.g., minimal indoor heating or AC) and/or upper bounds due to the capacities of the facilities, as well as transmission line capacities.

In order to solve the optimization problem \eqref{opt:woassump} using convex optimization methods, the following assumptions are made.
\begin{assumption}
       \label{assump:weak_util}
       All the utility functions $v^i_t(\cdot)$'s are twice differentiable and strictly concave.
\end{assumption}


\begin{assumption}
       \label{assump:feas_set}
       The feasible set $\cX$ is a polytope formed by several linear inequality constraints, and $\cX$ is coordinate convex, i.e., if $\ve{x} \in \cX$, then setting any of the components of $\ve{x}$ to 0 won't let it fall outside of set $\cX$.
\end{assumption}
By Assumption~\ref{assump:feas_set}, $\cX$ can be written as $\{\ve{x}|A\ve{x}\leq \ve{b}\}$ for some $A \in \Real^{L \times NT}$ and $\ve{b} \in \Real^{L}_+$, where $L$ is the number of linear constraints in $\cX$, and
\begin{align*}
       A &= \left[\ve{a}^1\ \ldots \ \ve{a}^L\right]^\Tr, \\
       \ve{a}^l &= \left[a^{1,l}_1\ \ldots\ a^{1,l}_T\ \ldots\ a^{N,l}_1\ \ldots\ a^{N,l}_T\right]^\Tr,\ l = 1,\ldots,L, \\
       \ve{b} &= \left[b^1,\ldots,b^L\right]^\Tr.
\end{align*}
The coordinate convexity in Assumption~\ref{assump:feas_set} is mainly used for the outside option required by the individual rationality. Under this assumption, for a feasible allocation $\ve{x}$, if any user $i$ changes her mind and chooses not to participate in the mechanism, the mechanism yields a feasible allocation with $\ve{x}^i=\ve{0}$ fixed.

With Assumptions~\ref{assump:weak_util},~\ref{assump:feas_set}, the energy community faces an optimization problem with a strictly concave and continuous objective function over a nonempty compact convex feasible set. Therefore, from convex optimization theory, the optimal solution for this problem always exists and should be unique\cite{polyak1987opt}.

Substituting the max function in \eqref{eq:cost} with a new variable $w$, the optimization problem in \eqref{opt:woassump} can be equivalently restated as
\begin{subequations}
       \label{opt:central}
       \begin{align}
              \underset{\ve{x},w}{\text{maximize}} \quad & \sum_{i=1}^N v^i(\ve{x}^i) - \sum_{t=1}^T p_t \sum_{i=1}^N x^i_t - p_0 w \\
              \label{cp2:feas}
              \text{subject to} \quad & A\ve{x} \leq \ve{b}, \\
              \label{cp2:peak}
              & \sum_{i=1}^N x^i_t \leq w, \ \forall t \in \left\{1,\ldots,T\right\}.
       \end{align}
\end{subequations}
The proof of this equivalency can be found in Appendix~\ref{appx:equiv_opt}.
The new optimization problem has a differentiable concave objective function with a convex feasible set, which means it is still a convex optimization, and therefore, KKT conditions are sufficient and necessary conditions for a solution $(\ve{x},\ve{\lambda},\ve{\mu})$ to be the optimal solution, where $\ve{\lambda}=[\lambda^1,\ldots,\lambda^L]^\Tr$ are the Lagrange multipliers for each linear constraint $\ve{a}^{l\Tr}\ve{x}\leq b^l$ in constraint $x\in\cX$, and $\ve{\mu}=[\mu_1,\ldots,\mu_T]^\Tr$ are the Lagrange multipliers for (\ref{cp2:peak}). The KKT conditions are listed as follows:

\begin{subequations}
\label{KKT}
\begin{enumerate}
       \item Primal Feasibility:
       \begin{align}
              \label{KKT:pfeas_x}
              \ve{x} &\in \cX,\\
              \label{KKT:pfeas_w}
              \sum_{i=1}^N x^i_t &\leq w.
       \end{align}
       \item Dual Feasibility:
       \begin{equation}
              \label{KKT:dual}
              \lambda^l \geq 0,\ l = 1, \ldots, L; \mu_t \geq 0,\ t = 1,\ldots,T.
       \end{equation}
       \item Complementary Slackness:
       \begin{align}
              \label{KKT:comp_lambda}
              \lambda^l (\ve{a}^{l\Tr} \ve{x} - b^l) &= 0,\ l = 1,\ldots,L, \\
              \label{KKT:comp_mu}
              \mu_t (\sum_{i=1}^N x^i_t - w) &= 0,\ t = 1,\ldots,T.
       \end{align}
       \item Stationarity:
       \begin{align}
              \label{KKT:stat_peak}
              p_0 &= \sum_t \mu_t, \\
              \label{KKT:stat_cons}
              \dot{v}^{i}_t(x^i_t) &= p_t + \sum_l \lambda^l a^{i,l}_t + \mu_t,\ t = 1,\ldots,T,\ i \in \cN.
       \end{align}
\end{enumerate}
\end{subequations}
where $\dot{v}^{i}_t(\cdot)$ is the first order derivative of $v^i_t(\cdot)$.

We conclude this section by pointing out once more that our objective is not to solve \eqref{opt:central} or \eqref{KKT} in a centralized or decentralized fashion. Such a methodology is well established and falls under the research area of centralized or decentralized (non-strategic) optimization. Furthermore, such a task can be accomplished only under the assumption that users
report their utilities (or related quantities, such as derivatives of utilities at specific points) truthfully, i.e., they do not act strategically.
Instead, our objective is to design a mechanism (i.e., messages and incentives) so that strategic users are presented with a game, the NE of which is designed so that it corresponds to the optimal solution of \eqref{opt:central} or \eqref{KKT}.

\subsection{Mechanism Design Preliminaries}

In an energy community, utilities are users' private information. Due to privacy and strategic concerns, users might not be willing to report their utilities. As a result, \eqref{opt:central} or \eqref{KKT} cannot be solved directly.
In order to solve \eqref{opt:central}, \eqref{KKT} under the settings stated above, we introduce a planner as an intermediary between the community and the energy company.
To incentivize users to provide necessary information for optimization, the planner signs a contract with users, which prespecifies the messages needed from users and rules for determining the allocation and taxes/subsidies from/to the users. The planner commits to the contract.
Informally speaking, the design of such contract is referred to as \emph{mechanism design}.

More formally, a \emph{mechanism} is a collection of message sets and an outcome function\cite{garg2008foundations}. Specifically, in resource allocation problems, a mechanism can be defined as a tuple $(\cM,\hat{x}(\cdot),\hat{t}(\cdot))$, where $\cM=\cM^1\times \ldots\times \cM^N$ is the space of message profile; $\hat{x}:\cM \mapsto \cX$ is an allocation function determining the allocation $\ve{x}$ according to the received message profile $m\in \cM$; and $\hat{t}: \cM \mapsto \Real^N$ is a tax function which defines the payments (or subsidies) of users based on $m$ (specifically, $\hat{t}=\{\hat{t}^i\}_{i\in \cN}$ with $\hat{t}^i:\cM \mapsto \Real$ defining the tax/subsidy function for user $i$). Once defined, the mechanism induces a game $(\cN,\cM,\{u^i\}_{i\in \cN})$. In this game, each user $i$ chooses her message $m^i$ from the message space $\cM^i$, with the objective to maximize her payoff $u^i(m)=v^i(\hat{x}^i(m)) - \hat{t}^i(m)$. The planner charges taxes and pays for the energy cost to the company, so the planner's payoff turns out to be $\sum_i \hat{t}^i(m) - J(\hat{x}(m))$ (the net income of the planner).

For the mechanism-induced game $\cG$, NE is an appropriate solution concept. At the equilibrium point $m^*$, if $\hat{x}(m^*)$ coincides with the optimal allocation (i.e., the solution of~\eqref{opt:central}), we say that the mechanism \emph{implements} the optimal allocation at $m^*$. A mechanism has the property of \emph{full implementation} if all the NE $m^*$'s implement the optimal allocation.

There are other desirable properties in a mechanism. \emph{Individual rationality} is the property that every one volunteers to participate in the mechanism-induced game instead of quitting.
For the planner, this means that the  sum of taxes $\sum_i \hat{t}^i(m^*)$ collected at NE is larger than the cost paid to the energy company $J(\hat{x}(m^*))$. In the context of this paper, \emph{strong budget balance} is the property that the sum of taxes is exactly the same as the
cost paid to the energy company, so no additional funds are required by the planner or the community to run the mechanism other than the true energy cost paid to the energy company.
In addition, if we use the solution concept of NE, one significant problem is how the users know the NE without full information. Therefore, some learning algorithm is needed to help users learn the NE. If under a specific class of learning algorithm, the message profile $m$ converges to NE $m^*$, then we say that the mechanism has learning guarantees with this certain class.


\section{The Baseline ``Centralized'' Mechanism}
\label{sec:centralized}

In this section we temporarily assume there are no communication constraints, i.e., all the message components are accessible for the calculations of the allocation and taxation. The mechanism designed under this assumption is called a ``centralized'' mechanism. In the next section we will extend this mechanism to an environment with communication constraints.

In the proposed centralized mechanism we define user $i$'s message $m^i$ as
\begin{equation*}
       m^i = \left(\left\{y^i_t\right\}_{t=1}^T, \left\{q^{i,l}\right\}_{l \in \cL},\left\{s^i_t\right\}_{t=1}^T, \left\{\beta^i_t\right\}_{t=1}^T\right).
\end{equation*}

Each message component above has an intuitive meaning. Message $y^i_t \in \Real$ can be regarded as the demand for time slot $t$ announced by user $i$.
Message $q^{i,l} \in \Real_+$ is the additional price that user $i$ expects to pay for the constraint $l$, which corresponds to the Lagrange multiplier $\lambda^l$.
Message $s^i_t \in \Real_+$ is proportional to the peak price that user $i$ expects to pay at time $t$. Intuitively, setting one $s^i_t$ greater than $s^i_{t'}$ means user $i$ thinks day $t$ is more likely to be the day with the peak demand rather than $t'$. This component corresponds to the Lagrange multiplier $\mu_t$.
Message $\beta^i_t \in \Real$ is the prediction of user $(i+1)$'s usage at time $t$ by user $i$. This message is included for technical reasons that will become clear in the following (for a user index $i \in \cN$, let $i-1$ and $i+1$ denote modulo $N$ operations).

Denote the message space of user $i$ by $\cM^i$, and the space of the message profile is represented as $\cM = \cM^1 \times \ldots \times \cM^N$. The allocation functions and the tax functions are functions defined on $\cM$. The allocation functions follow the simple definition:
\begin{equation}
       \label{def:alloc}
       \hat{x}^i_t(m) = y^i_t,\ t = 1,\ldots,T,\ \forall i \in \cN.
\end{equation}
i.e., users get exactly what they request.

Prior to the definition of the tax functions, we want to find some variable which acts like $\mu_t$ at NE. Although $s^i_t$ is designed to be proportional to $\mu_t$, it does not guarantee $\sum_t s^i_t = p_0$, which is KKT condition \eqref{KKT:stat_peak}.
To solve this problem, we utilize a technique similar to the proportional/radial allocation in \cite{kelly1998rate,yang2005revenue,basar2006,SiAn14b,SiAn17b} to shape the suggested peak price vector $\ve{s}$ into a form which satisfies \eqref{KKT:stat_peak}. 
For a generic $T$-dimensional peak price vector $\ve{\tilde{s}}=(\tilde{s}_1,\ldots,\tilde{s}_T)$ and a generic $T$-dimensional total demand vector $\ve{\tilde{y}}=(\tilde{y}_1,\ldots,\tilde{y}_T)$, define a radial pricing operator $\Rp^i: \Real^T_+ \times \Real^{T} \mapsto \Real^T_+$ as
\begin{subequations}
\begin{equation}
       \label{def:rp}
       \Rp^i (\ve{\tilde{s}},\ve{\tilde{y}}) = \left(\Rp^i_1(\ve{\tilde{s}},\ve{\tilde{y}}),\ldots,\Rp^i_T(\ve{\tilde{s}},\ve{\tilde{y}})\right),
\end{equation}
where
\begin{equation}
       \label{def:rpi}
       \Rp^i_t (\ve{\tilde{s}},\ve{\tilde{y}}) =
       \begin{cases}
            \frac{\tilde{s}_t}{\sum_{t'}\tilde{s}_{t'}}p_0, \quad \text{if }\ve{\tilde{s}}\neq \ve{0}, \\
              \frac{p_0 \cdot \one\{t\in\arg\underset{t'}{\max}\ \tilde{y}_{t'}\}}{\#(\arg\underset{t'}{\max}\ \tilde{y}_{t'})}, \quad \text{if }\ve{\tilde{s}}= \ve{0},
       \end{cases}
\end{equation}
\end{subequations}
and $\#(\arg\underset{t'}{\max}\ \tilde{y}_{t'})$ represents the number of elements in $\ve{\tilde{y}}$ that are equal to the maximum value.

%
The output of the radial pricing $\Rp(\cdot,\cdot)$ will be taken as the peak price in the subsequent tax functions. When the given suggested price vector $\ve{\tilde{s}}$ is a nonzero vector, the unit peak price will be allocated to each day proportional to $\tilde{s}_t$. If the suggested price vector $\ve{\tilde{s}}=\ve{0}$, then divide $p_0$ to the days with peak demand with equal proportion.

The tax functions are defined as
\begin{equation}
    \label{def:tax}
    \hat{t}^i(m) = \text{cost}^i(m) + \sum_{t=1}^T \text{pr$\boldsymbol{\beta}$}^{i}_t(m) + \sum_{l \in \cL} \text{con}^{i,l}(m) + \sum _{t=1}^T \text{con}^i_t(m),
\end{equation}
where
\begin{subequations}
\begin{align}
   \text{cost}^i(m) =& \sum_{t=1}^T (p_t+\Rp^i_t(\ve{s}^{-i},\ve{\zeta}^{-i})) \hat{x}^i_t(m) + \sum_{l \in \cL_i} q^{-i,l} \ve{a}^{i,l} \hat{\ve{x}}^i(m), \\
   \text{pr$\boldsymbol{\beta}$}^{i}_t(m) =& (\beta^i_t-y^{i+1}_t)^2, \\
    \text{con}^{i,l}(m) =& (q^{i,l} - q^{-i,l})^2 + q^{i,l} (b^l - \sum_{j \neq i}\ve{a}^{j,l} \ve{y}^j-\ve{a}^{i,l}\ve{\beta}^{i-1}), \label{tax1}\\
    \text{con}^i_t(m) =& (s^i_t-s^{-i}_t)^2 + s^i_t (z^{-i}-\zeta^{-i}_t),
\end{align}
\end{subequations}
and
\begin{subequations}
\begin{align}
       s^{-i}_t &= \frac{1}{N-1} \sum_{j\neq i} s^j_t \quad \forall i \ \forall t, \\
       q^{-i,l} &= \frac{1}{N-1} \sum_{j \neq i} q^{j,l} \quad \forall i \ \forall l,\\
       \zeta^{-i}_t &= \sum_{j\neq i} y^j_{t}+\beta^{i-1}_{t},\quad \forall i \ \forall t, \\
       z^{-i} &= \underset{t}{\max} \left\{\zeta^{-i}_t\right\} \quad \forall i,
\end{align}
\end{subequations}
and $\ve{a}^{i,l}$ is defined as $\ve{a}^{i,l}=[a^{i,l}_1,\ldots,a^{i,l}_T]$.

The tax function for user $i$ consists of three parts. 
The first part $\text{cost}^i(m)$ is the cost for the demand. According to this part, user $i$ pays the fixed price and the peak price for her demand. Note that the peak price at time $t$, $\Rp^i_t(\ve{s}^{-i},\ve{\zeta}^{-i})$, is generated by the vector of peak prices from all other agents, $\ve{s}^{-i}$, and the total demand from all other agents (agent $i$'s demand at time $t$ is approximated by $\beta^{i-1}_t$). As a result, the peak price is not controlled by user $i$ at all.
The second part $\text{pr}\boldsymbol{\beta}^i_t(m)$ ($\text{pr}\boldsymbol{\beta}$ stands for ``proxy-$\beta$'') is a penalty term for the imprecision of prediction $\ve{\beta}^i$, which incentivizes $\ve{\beta}^i$ to align with $\ve{y}^{i+1}$ at NE.
The third part consists of two penalty terms $\text{con}^{i,l}(m)$ and $\text{con}^i_t(m)$ for each constraint $l\in\cL$ and each peak demand inequality $t\in\{1,\ldots,T\}$, respectively. Both of them have a quadratic term that incentivizes consensus of the messages $q^{i,l}$ and $s^{i}_t$ among agents, respectively. In addition, they possess a form which looks similar to the complementary slackness conditions \eqref{KKT:comp_lambda}, \eqref{KKT:comp_mu}. This special design facilitates the suggested price to come to an agreement, and ensures the primal feasibility and complementary slackness hold at NE, which will be shown in Lemma \ref{lem:kkt1-3}.

The main property we want from this mechanism is full implementation. We expect the allocation scheme under the NE of the mechanism-induced game to coincide with that of the original optimization problem. Full implementation can be shown in two steps. First, we show that if there is a (pure strategy) NE, it must induce the optimal allocation. Then we prove the existence of such (pure strategy) NE.

From the form of the tax functions, we can immediately get the following lemma.

\begin{lemma}
       \label{lem:eqbeta}
       At any NE, for each user $i$,  the demand proxy $\beta^i_t$ is equal to the demand of her next neighbor, i.e., $\beta^i_t=y^{i+1}_t$ for all $t$.
\end{lemma}

\begin{IEEEproof}
       Suppose $m$ is a NE where there exists at least one user $i$, whose message $\ve{\beta}^i$ does not agree with next user's demand, i.e., $\ve{\beta}^i \neq \ve{y}^{i+1}$. Say, $\beta^i_t \neq y^{i+1}_t$ for some $t$. Then we can find a profitable deviation $\tilde{m}$, which keeps everything other than $\ve{\beta}^i$ the same as $m$, but modifies $\beta^i_t$ with $\tilde{\beta}^i_t=y^{i+1}_t$. Compare the payoff value $u_i$ before and after the deviation:
       \begin{equation*}
         \begin{split}
            u_i(\tilde{m}) - u_i(m) =& -(\tilde{\beta}^i_t-y^{i+1}_t)^2 + (\beta^i_t-y^{i+1}_t)^2 \\
            =& (\beta^i_t-y^{i+1}_t)^2 >0.
         \end{split}
       \end{equation*}
       Thus, if there is some $\ve{\beta}^i \neq \ve{y}^{i+1}$, user $i$ can always construct another announcement $\tilde{m}^i$, such that user $i$ get a better payoff.
\end{IEEEproof}

It can be seen from Lemma~\ref{lem:eqbeta} that the messages $\beta$ play an important role in the mechanism. 
They appear in two places in the tax functions. First, in the expression of $\zeta^{-i}_t = \sum_{j\neq i} y^j_{t}+\beta^{i-1}_{t}$ which is the total demand at time $t$ used in user $i$'s tax function. 
Second, in the expression for excess demand $b^l - \sum_{j \neq i}\ve{a}^{j,l} \ve{y}^j-\ve{a}^{i,l}\ve{\beta}^{i-1}$ for the $l$-th constraint.
Note that we do not want user $i$ to control these terms with her messages (specifically $y^i_t$) because she already controls her allocation directly and this will create technical difficulties. Indeed, quoting the self-announced demand in the tax function raises the possibility of unexpected strategic moves for user $i$ to obtain extra profit. Instead, using the proxy $\ve{\beta}^{i-1}$ instead of $\ve{y}^i$ eliminates user $i$'s control on his own slackness factor, while Lemma \ref{lem:eqbeta} guarantees that at NE these quantities become equal.

With the introduction of these proxies, we show in the following lemmas, that at NE, all KKT conditions required for the optimal solution are satisfied. First we prove primal feasibility (KKT 1) and complementary slackness (KKT 3) are ensured by the design of the penalty terms ``$\text{pr}$''s and constraint-related terms ``$\text{con}$''s, if we treat $\ve{q}$ and $\Rp(\ve{s},\ve{\zeta})$ as the Lagrange multipliers.

\begin{lemma}
       \label{lem:kkt1-3}
       At any NE, users' suggested prices are equal:
       \begin{align*}
              q^{i,l} &= q^l,\ \forall l \in \cL_i\ \forall i \in \cN,\\
              s^i_t &= s_t,\ t=1,\ldots,T,\ \forall i \in \cN.
       \end{align*}
       Furthermore, users' announced demand profile satisfies $\ve{y} \in \cX$, and the equal prices, together with the demand profile, have satisfy complementary slackness:
       \begin{align*}
              \ve{q}(A\ve{x} - \ve{b}) &= \ve{0}, \\
              s_t \left(z - \sum_i y^i_t\right) &= 0, \ \forall t = 1,\ldots,T,
       \end{align*}
       which implies
       \begin{equation}
              \label{comp:wrp}
              \Rp^i_t(\ve{s},\ve{\zeta}^{-i})\left(z - \sum_i y^i_t \right) = 0, \ \forall t = 1,\ldots,T,
       \end{equation}
       where $z$ is the peak demand during the billing period.
\end{lemma}
\begin{IEEEproof}
       The proof can be found in Appendix~\ref{appx:pf_lem:kkt1-3}.
\end{IEEEproof}

Dual feasibility (KKT 2) holds trivially by definition. We now show that stationarity condition  (KKT 4) holds at NE by imposing first order condition on the partial derivatives of user $i$'s utility w.r.t. their message components $y^i_t$'s.

\begin{lemma}
       \label{lem:stat}
       At NE, stationarity holds, i.e.,
       \begin{align}
              \label{eq:sta_price}
              \dot{v}^{i}_t(\hat{x}^i_t(m)) &= p_t + \Rp^i_t(\ve{s},\ve{\zeta}^{-i}) + \sum_{l \in \cL_i} q^l a^{i,l}_t,\\
              \label{eq:sta_peak}
              p_0 &= \sum_{t=1}^T \Rp^i_t(\ve{s},\ve{\zeta}^{-i}).
       \end{align}
\end{lemma}
\begin{IEEEproof}
       The proof is in Appendix \ref{appx:pf_lem:stat}.
\end{IEEEproof}

With Lemma \ref{lem:eqbeta}, \ref{lem:kkt1-3} and \ref{lem:stat}, it is straightforward to derive the first part of our result, i.e.,  efficiency of the allocation at any NE.

\begin{theorem}
       \label{thm:NEopt}
       For the mechanism-induced game $\cG$, if NE exist, then the NE result in the same allocation as the optimal solution to the centralized problem~\eqref{opt:central}.
\end{theorem}
\begin{IEEEproof}
       If $m^*$ is a NE, from Lemma~\ref{lem:eqbeta} and~\ref{lem:kkt1-3}, we know that at NE, $\ve{\beta}^{i*}=\ve{y}^{i+1}$, and all the prices $\ve{q}^{i*}$, $\ve{s}^{i*}$, and all the $\ve{\zeta}^{-i*}$ are the same among all the users $i\in\cN$. We denote these equal quantities by $\ve{y}^*,\ve{q}^*,\ve{s}^*$ and $\ve{\zeta}^*$.

       Consider the solution $sol=(\ve{x},w,\ve{\lambda},\ve{\mu})=(\ve{y}^*,\max_t \{\zeta^*_t\}, \ve{q}^*,\Rp(\ve{s}^*,\ve{\zeta}^*))$. From Lemma~\ref{lem:kkt1-3}, the solution $sol$ satisfies \eqref{KKT:pfeas_x},\eqref{KKT:pfeas_w},\eqref{KKT:comp_lambda} and \eqref{KKT:comp_mu} (primal feasibility and complementary slackness). From Lemma~\ref{lem:stat}, $sol$ has \eqref{KKT:stat_cons} and \eqref{KKT:stat_peak} (stationarity). The dual feasibility \eqref{KKT:dual} holds because of the nonnegativity of $\ve{q}$ and $\ve{s}$.

       Therefore, $sol$ satisfies all the four KKT conditions, which means the allocation $\hat{\ve{x}}(m^*)$ is the optimal allocation.
\end{IEEEproof}

The following theorem shows the existence of NE.
\begin{theorem}
       \label{thm:exist}
       For the mechanism-induced game $\cG$, there exists at least one NE.
\end{theorem}
\begin{IEEEproof}
       From the theory of convex optimization, we know that the optimal solution of \eqref{opt:central} exists. Based on this solution, one can construct a message profile which satisfies all the properties we present in Lemma \ref{lem:eqbeta},\ref{lem:kkt1-3},\ref{lem:stat} and prove there is no unilateral deviation for all users. The details are presented in Appendix \ref{appx:pf_thm:exist}.
\end{IEEEproof}

Full implementation indicates that if all users are willing to participate in the mechanism, the equilibrium outcome is nothing but the optimal allocation. For each user $i$, the payoff at NE will be
\begin{equation}
       \label{eq:final_payoff}
       \begin{split}
         u_i(m^*) &= v^i(\hat{\ve{x}}^i(m^*)) \\
         &- \sum_{t=1}^T \underbrace{\left(p_t + \Rp^i_t(\ve{s},\ve{\zeta}^{-i}) + \sum_{l\in \cL_i} q^{-i,l} a^{i,l}_t\right)}_{{\text{Aggregated unit price for } \hat{x}^i_t}} \hat{x}^i_t(m^*).
       \end{split}
\end{equation}
In other words, the users pay for their own demands by the aggregated unit prices given by the consensus at NE. By counting the planner as a participant of the mechanism with utility $\sum_{i\in\cN} \hat{t}^i(m^*)-J(\ve{x}^*)$, a strong budget balance is automatically achieved. However, there are still two questions remaining. Are the users willing to follow this mechanism or would they rather not participate? Will the planner have to pay extra money for implementing such mechanism? The two theorems below answer these questions.

\begin{theorem}[Individual Rationality for Users]
       \label{thm:ir}
       Assume agent $i$ gets $\ve{x}^i=\ve{0}$ and pays nothing if she chooses not to participate in the mechanism.
       Then, at NE, participating in the mechanism is weakly better than not participating, i.e.,
       \begin{equation*}
              u_i(m^*) \geq v^i(\ve{0}).
       \end{equation*}
\end{theorem}
\begin{IEEEproof}
       The main idea for the proof of Theorem~\ref{thm:ir} is to find a message profile with $m^{-i*}$, in which user $i$'s payoff is $v^i(\ve{0})$, and then we can argue that following NE won't be worse since $m^*$ is a best response to $m^{-i*}$. The details of the proof can be found in Appendix \ref{appx:pf_thm:ir}.
\end{IEEEproof}

\begin{theorem}[Individual Rationality for the Planner]
       \label{thm:bb}
       At NE, the planner does not need to pay extra money for the mechanism:
       \begin{equation}
              \label{eq:wbb}
              \sum_{i\in \cN} \hat{t}^i(m^*) - J(\hat{\ve{x}}(m^*)) \geq 0.
       \end{equation}
       Moreover, by a slight modification of the tax functions defined in~\eqref{def:tax}, the total payment of users and the energy cost achieve a balance at NE:
       \begin{equation}
              \label{eq:sbb}
              \sum_{i\in \cN} \tilde{t}^i(m^*) - J(\hat{\ve{x}}(m^*)) = 0.
       \end{equation}
\end{theorem}

\begin{IEEEproof}
       The verification of individual rationality of the planner can be done by substituting $m^*$ in \eqref{eq:wbb} directly. By redistributing the income of the planner back to the users in a certain way, the total payment of users is exactly $J(\hat{\ve{x}}(m^*))$ and consequently no money is left after paying the energy company. The details are left to Appendix \ref{appx:pf_thm:bb}.
\end{IEEEproof}

\section{Distributed Mechanism}
\label{sec:distributed}

In the previous mechanism, allocation functions and tax functions of users depend on the global message profile $m$. If one wants to compute the tax $\hat{t}^i$ for a certain user $i$, all messages $m^j$ for all $j\in \cN$ are needed. Such mechanisms are not desirable for environments with communication constraints, where such global message exchange is restricted. To tackle this problem, we provide a distributed mechanism, in which the calculation of the allocation and tax of a certain user depends only on the messages from the ``available'' users, and therefore satisfies the communication constraints. In this section, we will first introduce communication constraints using a message exchange network model. We then develop a distributed mechanism, which accommodates the communication constrains and preserves the desirable properties of the baseline centralized mechanism.

\subsection{Message Exchange Network}

In an environment with communication constraints, all the users are organized in a undirected graph $\Gr=(\cN,\cE)$, where the set of nodes $\cN$ is the set of users, and the set of edges $\cE$ indicates the accessibility to the message for each user. If $(i,j) \in \cE$, user $i$ can access the message of user $j$, i.e., the message of $j$ is available for user $i$ when computing the allocation and tax of user $i$, and vice versa. Here we state a mild requirement for the message exchange network:

\begin{assumption}
       The graph $\Gr$ is a connected graph.
\end{assumption}

In fact, the mechanism we are going to show will work for the cases where $\Gr$ is a tree. Although an undirected connected graph is not necessarily a tree, since we can always find a spanning tree from such graph, it is safe to consider the mechanism under the assumption that the given network has a tree structure. If that is not the case, the mechanism designer can claim a spanning tree from the original message exchange network, and design the mechanism only based on the tree instead of the whole graph (essentially some of the connections of the original graph will never be used for message exchanges).

The basic idea behind the decentralized modification of the baseline mechanism is intuitively straightforward. Looking at the tax function for user $i$ in the centralized mechanism we observe that several messages required are not coming from $i$'s immediate neighbors. For this reason we define new ``summary'' messages that are quoted by $i$'s neighbors and represent the missing messages. At the same time, for this to work, we add additional penalty terms that guarantee that the summary messages will indeed represent the needed terms at NE. 

Notice that in the previous mechanism, user $i$ is expected to announce a $\beta^i_t$ equal to the demand of the next user $(i+1)$, but here we might have $(i,i+1) \notin \cE$, and owing to the communication constraint, we are not able to compare $\beta^i_t$ with $y^{i+1}_t$. Instead, $\beta^i_t$ should be a proxy of the demand of user $i$'s direct neighbor. This motivates us to define the function $\phi(i)$, where $\phi(i) \in \cN(i)$, $\cN(i)$ is the set of user $i$'s neighbors (excluding $i$), and $\phi(i)=j$ denotes that in user $i$'s tax function, the proxy variable $\ve{\beta}$ used for user $i$'s $\text{con}_i^l(m)$ terms in her tax function is provided by user $j$. In other words, $\phi(i)$ is a ``helper'' for user $i$ who quotes a proxy of his demand whenever needed.

In the next part we are going to use the summaries of the demands to deal with the distributed issue. For the sake of convenience, we define $n(i,k)$ as the nearest user to user $k$ among the neighbors of user $i$ and user $i$ itself. $n(i,k)$ is well-defined because of the tree structure.
The proof is omitted here. The details can be found in\cite[Ch.~4, Sec.~7.1]{sinha2017mechanism2}.

\subsection{The Message Space}

\begin{figure}
       \centering
       \includegraphics[width=10cm]{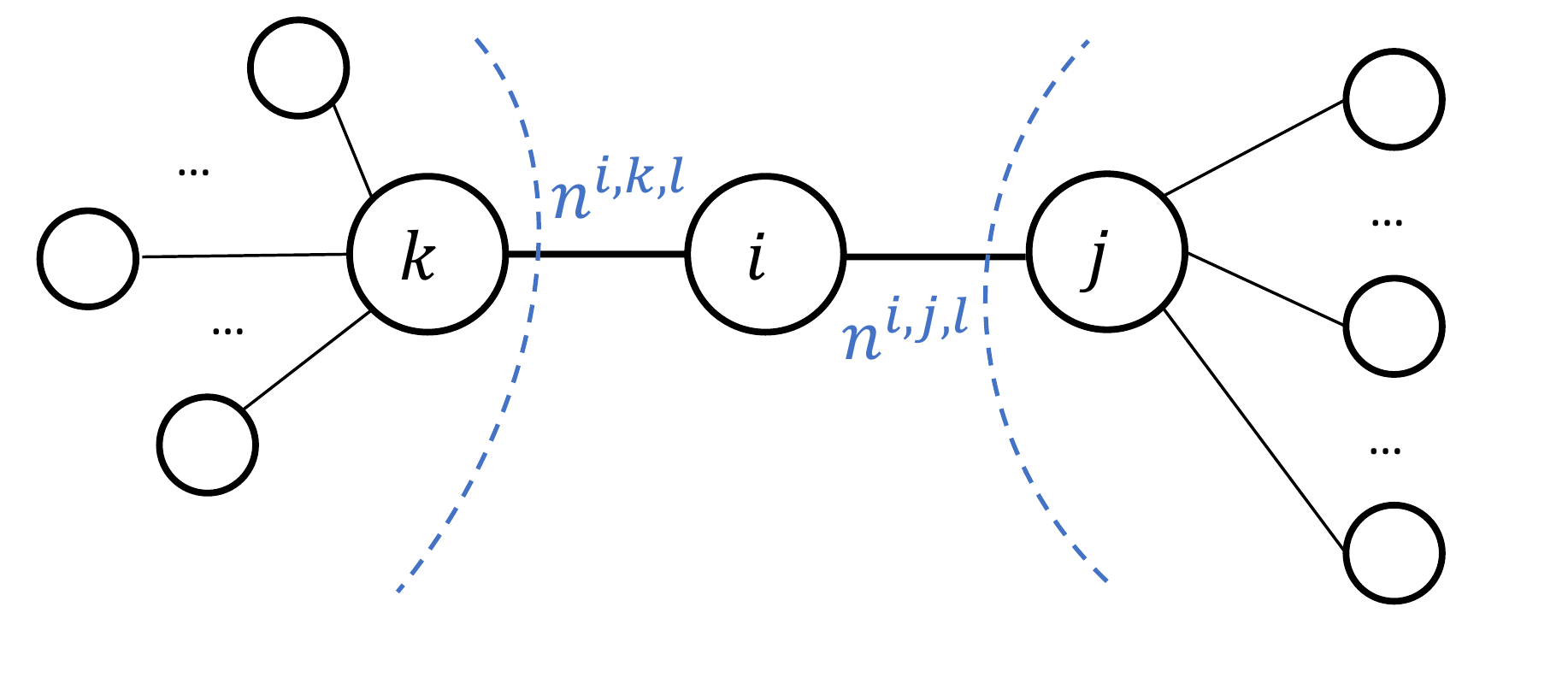}
       \caption{Proxies in the Message Exchange Network: for constraint~$l$, user~$i$ announces $n^{i,k,l}$ as a summary of demands for the tree on the left of $i$ (starting from $k$), and $n^{i,j,l}$ as a summary of demands for the tree on the right of $i$ (starting from $j$).\label{fig:proxy}}
\end{figure}

In the distributed mechanism, the message $m^i$ in user $i$'s message space $\cM^i$ is defined as
\begin{equation*}
       \begin{split}
              m^i = \left(\left\{y^i_t\right\}_{t=1}^T,\left\{q^{i,l}\right\}_{l\in\cL},\left\{s^i_t\right\}_{t=1}^T,\left\{\beta^{i,j}_t:\phi(j)=i\right\}_{t=1}^T,\right.\\
              \left.\left\{n^{i,j,l}:j\in \cN(i)\right\}_{l\in \cL},\left\{\nu^{i,j}_t:j\in \cN(i)\right\}_{t=1}^T\right).
       \end{split}
\end{equation*}
Here $n^{i,j,l}$ is a summary for demands of users related to constraint $l$ and connected to user $i$ via $j$ as depicted in Figure~\ref{fig:proxy}. Message $\nu^{i,j}_t$ serves a similar role for the peak demand. 

\subsection{The Allocation and Tax Functions}

The allocation functions $\hat{x}^i_t(m) = y^i_t$ are still straightforward. There are some modifications on tax functions, including adjustments on prices, consensus of new variables, and terms for complementary slackness.

\begin{equation}
       \label{def:disttax}
       \begin{split}
         \hat{t}^i(m) &= \text{cost}^i(m) + \sum\nolimits_l(\text{pr$\boldsymbol{n}$}^{i,l}(m) + \text{con}^{i,l}(m)) \\
         &+ \sum\nolimits_t (\text{pr$\boldsymbol{\beta}$}^i_t(m) + \text{pr$\boldsymbol{\nu}$}^i_t(m)+ \text{con}^i_t(m)),
       \end{split}
   \end{equation}
   where
   \begin{subequations}
   \begin{align}
       \text{cost}^i(m) &= \sum_{t=1}^T (p_t+\Rp^{-i}_t(\ve{s}^{-i},\ve{\zeta}^{-i})) \hat{x}^i_t(m) + \sum_{l \in \cL_i} q^{-i,l} \ve{a}^{i,l} \hat{\ve{x}}^i(m), \\
       \text{con}^{i,l}(m) &= (q^{i,l}-q^{-i,l})^2 + q^{i,l} (b^l-\sum_{j\in \cN(i)}f^{i,j,l}-\ve{a}^{i,l}\ve{\beta}^{\phi(i),i}),  \label{tax2}\\
       \text{con}^i_t(m) &= (s^i_t-s^{-i}_t)^2 + s^i_t (z^{-i}-\zeta^{-i}_t), \\
       \text{pr$\boldsymbol{n}$}^{i,l}(m) &= \sum_{j\in\cN(i)} \left(n^{i,j,l}-f^{i,j,l}\right)^2, \\
       \text{pr$\boldsymbol{\beta}$}^i_t(m) &=\sum_{j:\phi(j)=i} (\beta^{i,j}_t-y^j_t)^2, \\
       \text{pr$\boldsymbol{\nu}$}^i_t(m) &= \sum_{j\in\cN(i)}\left(\nu^{i,j}_t - f^{i,j}_t\right)^2 \\
       f^{i,j,l} &= \ve{a}^{j,l}\ve{y}^j + \sum_{h\in\cN(j)\backslash\{i\}}n^{j,h,l}, \\
       f^{i,j}_t &= y^j_t+\sum_{h\in\cN(j)\backslash\{i\}}\nu^{j,h}_t.
   \end{align}
   \end{subequations}
and
\begin{subequations}
\begin{align}
       s^{-i}_t &= \frac{1}{|\cN(i)|} \sum_{j\in \cN(i)} s^j_t \quad \forall i \ \forall t, \\
       q^{-i,l} &= \frac{1}{|\cN(i)|} \sum_{j \in \cN(i)} q^{j,l} \quad \forall i \ \forall l,\\
       \zeta^{-i}_t &= \sum_{j\in\cN(i)}f^{i,j}_t+\beta^{\phi(i),i}_t,\quad \forall i \ \forall t,\\
       z^{-i} &= \underset{t}{\max} \left\{\zeta^{-i}_t\right\} \quad \forall i.
\end{align}
\end{subequations}

In order to see intuitively how the decentralized mechanism works take as an example the term $\text{con}^{i,l}(m)$ in \eqref{tax2} which is a modified version of \eqref{tax1} repeated here for convenience
$\text{con}^{i,l}(m) = (q^{i,l} - q^{-i,l})^2 + q^{i,l} (b^l - \sum_{j \neq i}\ve{a}^{j,l} \ve{y}^j-\ve{a}^{i,l}\ve{\beta}^{i-1})$ related to the $l$-th constraint.
Other than the quadratic term which is identical in both expressions, the difference between the centralized and decentralized versions is in the expression $\sum_{j \neq i}\ve{a}^{j,l} \ve{y}^j + \ve{a}^{i,l}\ve{\beta}^{i-1}$ and $\sum_{j\in \cN(i)}f^{i,j,l}+\ve{a}^{i,l}\ve{\beta}^{\phi(i),i}$, respectively. The second term in each of these expressions relates to the proxy $\ve{\beta}^{i-1}$ which in the decentralized version is substituted by the proxy $\ve{\beta}^{\phi(i),i}$ due to the fact that the proxy for $y^i$ is not provided by user $i-1$ anymore but is provided by user $i$'s helper $\phi(i)$.
The first term, $\sum_{j \neq i}\ve{a}^{j,l} \ve{y}^j=\sum_{j \in\cN(i)}\ve{a}^{j,l} \ve{y}^j+\sum_{j \notin\cN(i)\cup \{i\}}\ve{a}^{j,l} \ve{y}^j$, which cannot be directly evaluated in the decentralized version (since it depends on messages outside the neighborhood of $i$) is now evaluated as $\sum_{j\in \cN(i)}f^{i,j,l}=\sum_{j\in \cN(i)} \ve{a}^{j,l}\ve{y}^j + \sum_{j\in \cN(i)}\sum_{h\in\cN(j)\backslash\{i\}}n^{j,h,l}$. It should now be clear that the role of the new messages $n^{j,h,l}$ quoted by the neighbors $j\in\cN(i)$ of $i$, is to summarize the total demands of other users. Furthermore, the additional quadratic penalty terms will have to effectuate this equality. This idea is made precise in the next section.

\subsection{Properties}

It is clear that this mechanism is distributed, since all the messages needed for the allocation and tax functions of user $i$ come from her neighborhood $\cN(i)$ and herself. Due to way the messages and taxes are designed, the proposed mechanism satisfies properties similar to those in Lemma \ref{lem:kkt1-3}, \ref{lem:stat}, and consequently Theorem \ref{thm:NEopt}, \ref{thm:exist}. The reason is that the components $n$ and $\nu$ behave the same as the absent $\ve{y}^h,\ h \notin \cN(i)$ in  user $i$'s functions at NE, which makes the proofs of the properties in previous mechanism still work here. We elaborate on these properties in the following.

\begin{lemma}
       \label{lem:dist_eqbeta}
       At any NE, we have the following results regarding the proxy messages:
       \begin{align}
              \label{eqprx:beta}
              \beta^{i,j}_t=&y^j_t,& \forall j: \phi(j)=i, \\
              \label{eqprx:n}
              n^{i,j,l}=& \ve{a}^{j,l} \ve{y}^j + \sum_{h \in \cN(j)\backslash\{i\}}n^{j,h,l},& \forall i,\forall j\in\cN(i),\forall l\in \cL,\\
              \label{eqprx:nu}
              \nu^{i,j}_t=&y^j_t+\sum_{h\in\cN(j)\backslash\{i\}} \nu^{j,h}_t,&\forall t, \forall i, \forall j \in \cN(i).
       \end{align}
\end{lemma}
\begin{IEEEproof}
       $\beta^{i,j}, n^{i,j,l}$ and $\nu^{i,j}_t$ only appear in the quadratic penalty terms of user $i$'s tax function. Therefore, for any user $i$, the only choice to minimize the tax is to bid $\beta^{i,j}, n^{i,j,l}$ and $\nu^{i,j}_t$ by \eqref{eqprx:beta}-\eqref{eqprx:nu}.
\end{IEEEproof}

Now, based on the structure of the message exchange network, we have

\begin{lemma}
       \label{lem:mes_comm}
       At any NE, $n^{i,j,l}$ and $\nu^{i,j}_t$ satisfy
       \begin{align}
              \label{eqprx:n:ctr}
              n^{i,j,l}=&\sum_{t=1}^T  \sum_{h:n(i,h)=j} a^{h,l}_t y^{h}_t,\qquad \forall i \in \cN,\forall j\in \cN(i), \forall l\in \cL,\\
              \label{eqprx:nu:ctr}
              \nu^{i,j}_t=&\sum_{h:n(i,h)=j}y^h_t,\forall i \in \cN,\qquad\forall j\in\cN(i), \forall t\in \cT.
       \end{align}
\end{lemma}
\begin{IEEEproof}
       The proof is presented in Appendix \ref{appx:pf_lem:mes_comm}.
\end{IEEEproof}

With Lemma~\ref{lem:mes_comm}, we immediately obtain the following results.

\begin{lemma}
       \label{lem:dist_sum}
       At any NE, for all user $i$, we have
       \begin{align}
              \label{eqprx:totn}
              \sum_{j \in \cN(i)} f^{i,j,l} =& \sum_{j \in \cN\backslash\{i\}} \ve{a}^{j,l}\ve{y}^j,\quad \forall l \in \cL_i,\\
              \label{eqprx:totnu}
              \sum_{j \in \cN(i)} f^{i,j}_t = & \sum_{j \in \cN\backslash\{i\}} y^j_t, \quad \forall t\in \cT.
       \end{align}
\end{lemma}
\begin{IEEEproof}
       At NE, by directly substituting:
       \begin{equation*}
              \begin{split}
                     \sum_{j \in \cN(i)} f^{i,j,l} &= \sum_{j \in \cN(i)} \left(\ve{a}^{j,l}\ve{y}^j+\sum_{h\in\cN(j)\backslash\{i\}}n^{j,h,l}\right) \\
                     &=  \sum_{j \in \cN(i)} \left(\ve{a}^{j,l}\ve{y}^j+\sum_{h\in\cN(j)\backslash\{i\}}  \sum_{k:n(j,k)=h} \ve{a}^{k,l} \ve{y}^k  \right) \\
                     &= \sum_{j \in \cN(i)} \left( \sum_{h:n(i,h)=j}\ve{a}^{h,l}\ve{y}^h \right) = \sum_{j \in \cN\backslash\{i\}} \ve{a}^{j,l}\ve{y}^j,\quad \forall l \in \cL_i,
              \end{split}
       \end{equation*}
       The third equality holds by the fact that the users in set $\{k|h\in \cN(j) \backslash \{i\}, n(j,k) = h\}$ are the ones that are not in the subtree starting from a single branch $(j,i)$ with root $j$, which is exactly $\{k|n(i,k)=j\} \backslash \{j\}$.
       
       \eqref{eqprx:totnu} holds for the similar reason.
\end{IEEEproof}

Lemma~\ref{lem:dist_sum} plays a similar role as Lemma~\ref{lem:eqbeta}. With Lemma~\ref{lem:dist_sum}, the properties in Lemma~\ref{lem:kkt1-3} and~\ref{lem:stat} can be reproduced in the distributed mechanism. We then obtain the following theorem.

\begin{theorem}
       \label{thm:fulimp-dist}
       For the mechanism-induced game $\cG$, NE exist. Furthermore, any NE of game $\cG$ induces the optimal allocation.
\end{theorem}
\begin{IEEEproof}
       By substituting \eqref{eqprx:totn}, \eqref{eqprx:totnu} in \eqref{def:disttax}, we obtain exactly the same form of the tax function in centralized mechanism on equilibrium, which yields the desirable results as shown in Lemmas~\ref{lem:kkt1-3}, \ref{lem:stat}. We conclude that any NE induces the optimal allocation. The existence of NE can be proved by a construction similar to that of Theorem \ref{thm:exist}.
\end{IEEEproof}

As was true in the baseline centralized mechanism, in the distributed case, the planner may also have the concerns whether the users have incentive to participate, and whether the mechanism requires external sources of funds to maintain the balance. As it turns out, Theorems \ref{thm:ir} and \ref{thm:bb} still hold here. As a result, the users are better off joining the mechanism, and the market has a balanced budget. The proofs and the construction of the subsidies can be done in a manner similar to the centralized case and therefore are omitted.

\section{A Learning Algorithm for the Centralized Mechanism}
\label{sec:learn}

The property of full implementation ensures that social welfare maximization can be reached if all the participants reach NE in the mechanism-induced game, and no one could get unilateral profitable deviation at NE. Nevertheless, it is troublesome for participants to anticipate NE as the outcome if none of them knows (or can calculate) NE without knowledge of other users' utilities. To settle this issue, one can design a learning algorithm to help participants learn the NE in an online fashion. In this section, we present such a learning algorithm for the centralized mechanism discussed in Section \ref{sec:centralized}. Instead of using Assumption \ref{assump:weak_util}, here we make a stronger assumption in order to obtain a convergent algorithm.

\begin{assumption}
       \label{assump:util}
       All the utility functions $v^i_t(\cdot)$'s are proper, twice differentiable concave functions with $\delta$-strong concavity.
\end{assumption}
Here $\delta$-strong concavity of a function $g(\cdot)$ is defined by the $\delta$-strong convexity of $-g(\cdot)$. A function $f(\cdot)$ is strongly convex with parameter $\delta$ if
\begin{equation*}
       f(\ve{y}) \geq f(\ve{x}) + \nabla f(\ve{x})^\Tr (\ve{y} - \ve{x})+ \frac{\delta}{2} ||\ve{y}-\ve{x}||^2.
\end{equation*}

The design of the learning algorithm involves three steps. First, we find the relation between NE and the optimal solution of the original optimization problem. This step has been done in the proof of Theorem~\ref{thm:NEopt}: we see in NE, $\ve{y}^*$ coincides with $\ve{x}^*$ in the optimal allocation, and $\ve{q}^{i*}$ equals $\ve{\lambda}^*$, and the components of $\ve{s}^{i*}$ are proportional to the components of $\ve{\mu}^*$. Then, by Slater's condition, strong duality holds here, so we connect the Lagrange multipliers $\ve{\lambda}^*,\ve{\mu}^*$ with the optimal solution of the dual problem. Due to the strong concavity of the utilities and stationarity, given $\ve{\lambda}^*$ and $\ve{\mu}^*$, the optimal allocation $\ve{x}^*$ can be uniquely determined. Finally, if we can find an algorithm to solve the dual problem, the design is done.

The first two steps are straightforward. For the third one, we can see the dual problem is also a convex optimization problem, so projected gradient descent (PGD) is one of the choices for the learning algorithm. The proof of convergence of PGD is not trivial. In the proof developed in \cite{polyak1987opt}, the convergence of PGD holds when (a) the objective function is $\beta$-smooth and (b) the feasible set is closed and convex. In Appendix~\ref{appx:pf_conv} we show that (a) is satisfied by Assumption \ref{assump:util}.
To check (b), we need to find  a feasible set for the dual variables. Since in PGD of the dual problem, the gradient of the dual function turns out to be a combination of functions of the form $(\dot{v}^i_t)^{-1}(\cdot)$, the feasible set should satisfy two requirements: first, all the elements are in the domain of the dual function's gradient in order to make every iteration valid; second, $(\lambda^*,\mu^*)$ is in the feasible set so that we won't miss it. With these requirements in mind, we make Assumption \ref{assump:price_set} and construct a feasible set for the dual problem based on that.
\begin{assumption}
       \label{assump:price_set}
       For each utility $v^i_t(\cdot)$, there exist $\underline{r}^i_t,\bar{r}^i_t \in \Real$ satisfying
       \begin{enumerate}
              \item $\forall \ve{x} \in \cX$, $\dot{v}^{i}_t(x^i_t) \in [\underline{r}^i_t,\bar{r}^i_t]$,
              \item $\forall p \in [\underline{r}^i_t,\bar{r}^i_t]$,  $\exists x^i_t \in \Real, \text{s.t. } \dot{v}^{i}_t(x^i_t)=p$.
       \end{enumerate}
\end{assumption}

Before we explain this assumption, define:

 $\underline{\ve{r}}=[\underline{r}^1_1, \ldots, \underline{r}^1_T, \ldots, \underline{r}^N_T]^\Tr$, $\bar{\ve{r}}=[\bar{r}^1_1, \ldots, \bar{r}^1_T, \ldots, \bar{r}^N_T]^\Tr$,

 $\ve{p}=[p_1 \ldots p_T]^\Tr$, $\tilde{\ve{p}} = \ve{1}_N \otimes \ve{p}$ and
\begin{equation}
       \label{def:tilde}
       \tilde{A} = \left(
       \begin{array}{c}
              A\\
              \ve{1}_{N}^\Tr \otimes I_T
       \end{array}
       \right), \
       \tilde{\ve{\lambda}} = \left(
              \begin{array}{c}
                     \ve{\lambda} \\
                     \ve{\mu}
              \end{array}
       \right), \
       \tilde{\ve{p}} = \ve{1}_N \otimes \ve{p}.
\end{equation}
where $\otimes$ represents Kronecker product of matrices. Then define a set of proper prices $\cP$ as the feasible set for the dual problem:
\begin{equation*}
       \cP = \{\tilde{\ve{\lambda}}=(\ve{\lambda},\ve{\mu}) \geq 0: \underline{\ve{r}} \leq \tilde{A}^\Tr \tilde{\ve{\lambda}} + \tilde{\ve{p}} \leq \bar{\ve{r}},\ \ve{1}_T^\Tr \ve{\mu} = p_0\}.
\end{equation*}
Observe that by stationarity, the $((i-1)T+t)$-th entry of $\tilde{A}^\Tr \tilde{\ve{\lambda}} + \tilde{\ve{p}}$ equals $\dot{v}^i_t(x^{i*}_t)$ in optimal solution. Consequently, Assumption \ref{assump:price_set} implies two things: first, $\tilde{\ve{\lambda}}^* \in \cP$; second, all $\tilde{A}^\Tr \tilde{\ve{\lambda}} + \tilde{\ve{p}}$ can be a vector of $\dot{v}^i_t$'s on some $\ve{x}\in \Real^{NT}$ if $\tilde{\ve{\lambda}} \in \cP$. Hence, with Assumption \ref{assump:price_set}, it is safe to narrow down the feasible set of the dual problem to $\cP$ without changing the optimal solution. Furthermore, for all the price vectors in $\cP$, $(\dot{v}^i_t)^{-1}(\cdot)$ in PGD can be evaluated. Back to condition (b) stated above, since $\cP$ is closed and convex, PGD is convergent in this case.

Based on all the assumptions and the PGD method, we propose Algorithm~\ref{algo:dyn} as a learning algorithm for the NE of the centralized mechanism.

\begin{algorithm}[H]
       \caption{The learning algorithm for the centralized mechanism}
       \label{algo:dyn}
       \LinesNumbered
       \KwData{Time index $k$, a set of proper prices $\cP$, a vector of initial prices $(\ve{q}^0$, $\ve{s}^0) \in \cP$, message profiles $m(k)$, iteration step size $\alpha$, number of iterations $K$.}
       \KwResult{Message profile $m(K)=(\ve{y}(K),\ve{\beta}(K),\ve{q}(K),\ve{s}(K))$.}
       $k=0$, $q^{i,l}(0)=q^{0,l}$, $s^i_t(0)=s^0_t,\quad \forall i\in \cN,l\in \cL,t\in \cT$\;
       $y^i_t(0) = (\dot{v}^{i}_t)^{-1}(p_t+\sum_{l\in \cL}a^{i,l} q^{i,l}(0)+s^i_t(0)),\quad \forall i \in \cN, t\in \cT$\;
       \While{$k<K$ and $||m(k)-m(k-1)||>0$}{
              $\tilde{q}^{i,l}(k+1) = q^{i,l}(k) - \alpha (b^l - \ve{a}^l \ve{y}(k)), \quad \forall i \in \cN, l \in \cL$\;
              $\tilde{s}^i_t(k+1) = s^i_t(k) + \alpha\sum_j y^j_t(k), \quad \forall i \in \cN, t \in \cT$\;
              $(\ve{q}^i(k+1),\ve{s}^i(k+1)) = \mathbf{Proj}_\cP \left(\tilde{\ve{q}}^i(k+1),\tilde{\ve{s}}^i(k+1)\right), \quad \forall i \in \cN$\;
              $y^i_t(k+1) = (\dot{v}^{i}_t)^{-1}(p_t+\sum_{l\in \cL}a^{i,l}_t q^{i,l}(k+1)+s^i_t(k+1)),\quad \forall i \in \cN, t\in \cT$\;
              $k \leftarrow k+1 $\;
       }
       $\beta^{i}_t(K) = y^{i+1}_t(K)\quad \forall i \in \cN, t \in \cT$\;
\end{algorithm}

The convergence of PGD yields the convergence of proposed learning algorithm:
\begin{theorem}
       \label{thm:conv}
       Choose a step size $\alpha \leq \delta'/\lVert A \rVert$, where $\lVert A \rVert$ is $A$'s spectral norm, $\delta'$ is the parameter of strong concavity of the centralized objective function. As the number of iterations $K$ grows, the distance between the computed price vector $(\ve{q}(K),\ve{s}(K))$ and the optimal price vector $(\ve{q}^*,\ve{s}^*)$ is non-increasing. Furthermore, $\lim_{K\to\infty}m(K) = m^*$, where $m^*$ is the NE.
\end{theorem}
\begin{IEEEproof}
       See Appendix \ref{appx:pf_conv}.
\end{IEEEproof}


\section{A Concrete Example}

To give a sense of how the two mechanisms and the learning algorithm work, we provide a simple non-trivial example here. We will first present the original centralized problem for the example, and then identify the NE of the centralized mechanism based on the properties we found. For the distributed mechanism, we will illustrate how the proxy variables at NE are determined with a simple example of a message exchange network. Lastly, we implement the learning algorithm for the centralized mechanism.

\subsection{The Demand Management Optimization Problem}

In the energy community, assume there are three users in the user set $\cN = \{1,2,3\}$, and $T=2$ days in a billing period. Suppose user $i$ on day $t$ has the following utility function:
\begin{equation*}
       v^i_t(x^i_t) = i \cdot t \cdot \ln (2 + x^i_t) .
\end{equation*}
Set $p_1 = 0.1$, $p_2 = 0.2$, and the peak price $p_0 = 0.05$. We adopt the following centralized problem as a concrete example:
\begin{subequations}
       \begin{align*}
              \underset{\ve{x}}{\text{maximize}} \quad & \sum_{t=1}^2 \sum_{i=1}^3 i \cdot t \cdot \ln (2 + x^i_t)  - J(\ve{x}) \\
              \text{subject to} \quad & x^i_t \geq -1, \ i=1,2,3, \ t = 1, 2, \\
              & \sum_{t=1}^2 (x^1_t + x^2_t + x^3_t) \leq 2,
       \end{align*}
\end{subequations}
where $J(\ve{x}) = 0.1 \cdot \sum_{t=1}^2 t \cdot \left( \sum_{i=1}^3 x^i_t \right) + 0.05 \cdot \max_t\{\sum_{i=1}^3 x^i_t\}$.

The solution to this problem is approximately\footnote{The exact solution is $x^{1*}_1=-1$, $\lambda^{7*}=(249+\sqrt{106201})/520$, $\mu_2=0.05$, and $x^{i*}_t=2/(\lambda^{7*}+p_t+\mu_t^*)-2$ for $(i,t)\neq (1,1)$, $\lambda^{1*}=\lambda^{7*}+p_1-1/(x^{1*}_1+2)$. $\lambda^{l*}=0$ for $l=2,\ldots,6$, $\mu_1^*=0$. The interested readers can verify it by using KKT conditions.}
\begin{equation*}
       (x^{1*}_1,x^{1*}_2,x^{2*}_1,x^{2*}_2,x^{3*}_1,x^{3*}_2)=(-1.0000, -0.5246, -0.3410,0.9508,0.4885,2.4263).
\end{equation*}
The lower bound constraint for $x^1_1$ and the upper bound constraint for the sum are active. Thus, according to KKT conditions, $\lambda^{l*} = 0$ for $l=2,\ldots,6$, and $\lambda^{1*} = 0.2056$, $\lambda^{7*} = 1.1056$ by stationarity. The total demands of Day 1 and Day 2 are $-0.8525$ and $2.8525$ respectively, so Day 2 has the peak demand $w^*=2.8525$, Day 1 charges no peak price ($\mu_1^*=0$), and Day 2 has an extra unit peak price $\mu_2^*=0.05$.

\subsection{The Centralized Mechanism}

For this example, in the centralized mechanism, user $i$ needs to choose her message $m^i$ with the following components:
\begin{equation*}
       m^i = (y^i_1, y^i_2, \{q^{i,l}\}_{l=1}^7, s^i_1,s^i_2, \beta^i_1,\beta^i_2).
\end{equation*}

For the sake of brevity, let's take user~1 for example. In this problem setting, user~1 needs to report her demands for two days ($y^1_1,y^1_2$), suggest a set of prices for constraints~1-7 ($q^{1,1},\ldots,q^{1,7}$), suggest unit peak prices for two days (quantities $s^1_1,s^1_2$ do not necessarily sum up to $p_0=0.05$), and lastly, provide proxies $\beta^1_1,\beta^1_2$ for user~2's demands.

User~1's tax function is
\begin{equation*}
       \begin{split}
              \hat{t}^1(m) &= \sum_{t=1}^2 \left(p_t+\Rp^1_t(s^{-1},\zeta^{-1})\right)y^1_t - q^{-1,1} y^1_1 - q^{-1,2} y^1_2 + q^{-1,7} (y^1_1 + y^1_2) \\
              &+ \sum_{t=1}^2(\beta^1_t-y^2_t)^2 + (q^{1,1} - q^{-1,1})^2 + q^{1,1} (1 + \beta^3_1) + (q^{1,2} - q^{-1,2})^2 +q^{1,2} (1 + \beta^3_2) \\
              &+ \sum_{l=3}^6 (q^{1,l} - q^{-1,l})^2 + q^{1,3} (1+y^2_1) + q^{1,4} (1+y^2_2) + q^{1,5} (1+y^3_1) + q^{1,6} (1+y^3_2)\\
              &+(q^{1,7} - q^{-1,7})^2 + q^{1,7} (2-y^2_1-y^2_2-y^3_1-y^3_2-\beta^3_1-\beta^3_2) \\
              &+ (s^1_1 - s^{-1}_1)^2 + s^1_1 (z^{-1}-\zeta^{-1}_1) + (s^1_2 - s^{-1}_2)^2 + s^1_2 (z^{-1}-\zeta^{-1}_2)
       \end{split}
\end{equation*}
where
\begin{align*}
       q^{-1,l} &= (q^{2,l}+q^{3,l}) / 2, \\
       s^{-1}_t &= (s^2_t + s^3_t) / 2, \\
       \zeta^{-1}_t &= \beta^3_t + y^2_t + y^3_t, \\
       z^{-1} &= \max \left\{ \zeta^{-1}_1, \zeta^{-1}_2 \right\},
\end{align*}
and according to the definition \eqref{def:rp}, \eqref{def:rpi} of the radial pricing operator, we have
\begin{equation*}
       \Rp^1_t(\ve{s}^{-1},\zeta^{-1}) = \begin{cases}
              \frac{s^{-1}_t}{s^{-1}_1+s^{-1}_2} p_0, & \text{if } s^{-1}_1 + s^{-1}_2 > 0,\\
              p_0, & \text{if } s^{-1}_1=s^{-1}_2 = 0 \text{ and }\zeta^{-1}_t > \zeta^{-1}_{t'} (t' \neq t), \\
              p_0/2, & \text{if } s^{-1}_1=s^{-1}_2 = 0 \text{ and }\zeta^{-1}_t = \zeta^{-1}_{t'} (t' \neq t), \\
              0, & \text{if } s^{-1}_1=s^{-1}_2 = 0 \text{ and }\zeta^{-1}_t < \zeta^{-1}_{t'} (t' \neq t).
       \end{cases}
\end{equation*}

From Theorem \ref{thm:NEopt} we know that at NE, user 1's message $m^{*1}$ is such that $\ve{y}^1$ corresponds to the optimal solution $x^1$, $\ve{q}^1$ equals the optimal Lagrange multiplier $\ve{\lambda}$, $\ve{\beta}^1$ equals $\ve{y}^2$, and finally 
$\ve{s}^1$ is proportional to the Lagrange multiplier $\ve{\mu}$.


\subsection{The Distributed Mechanism}

In this subsection we will first demonstrate the modifications on message spaces compared to the centralized mechanism, and then show how the newly introduced components $n$ and $\nu$ work. The specific NE can be determined in a similar way with that of the centralized mechanism and therefore omitted.

\begin{figure}
       \centering
       \includegraphics[width=5cm]{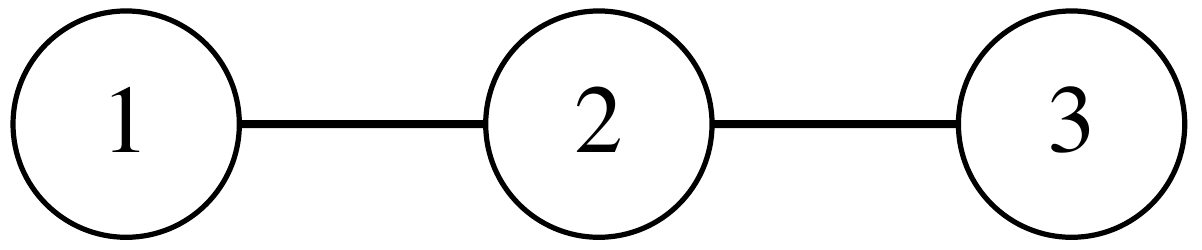}
       \caption{Message exchange network: Users~1 and~2, users~2 and~3 are neighbors respectively. User~1's message is invisible to user~3, and vice versa.\label{fig:mes_ex_net}}
\end{figure}

Assume the energy community has communication constraints with the message exchange network depicted in Figure~\ref{fig:mes_ex_net}. Apart from the network topology, the $\phi$-relation which indicates the responsibility of proxy $\ve{\beta}$ is also an important part of distributed mechanism. Here we set $\phi(1)=2, \phi(2)=1,\phi(3)=2$.
Then for proxy variables $\ve{\beta}^{\phi(i),i}, i=1,2,3$, $\ve{\beta}^{2,1}$ in user~1's tax is provided by user~2, $\ve{\beta}^{1,2}$ in user~2's tax is provided by user~1, and $\ve{\beta}^{2,3}$ in user~3's tax is provided by user~2.

For this message exchange network, the message components for each user are
\begin{align*}
       m^1 &= \left(y^1_1,y^1_2,\{q^{1,l}\}_{l=1}^7, \{s^1_t\}_{t=1}^2,\{ \beta^{1,2}_t\}_{t=1}^2,\{n^{1,2,l}\}_{l=1}^7,\{\nu^{1,2}\}_{t=1}^2 \right), \\
       m^2 &= \left(y^2_1,y^2_2,\{q^{2,l}\}_{l=1}^7, \{s^2_t\}_{t=1}^2, \{ \beta^{2,1}_t\}_{t=1}^2, \{ \beta^{2,3}_t\}_{t=1}^2,\{n^{2,1,l}\}_{l=1}^7,\{n^{2,3,l}\}_{l=1}^7,\{\nu^{2,1}\}_{t=1}^2,\{\nu^{2,3}\}_{t=1}^2 \right), \\
       m^3 &= \left(y^3_1, y^3_2, \{q^{3,l}\}_{l=1}^7, \{s^3_t\}_{t=1}^2,\{n^{3,2,l}\}_{l=1}^7,\{\nu^{3,2}\}_{t=1}^2\right).
\end{align*}
Therefore, in the distributed mechanism, users are still required to provide their demands $\ve{y}$, suggested unit prices $\ve{q}$ and suggested unit peak prices $\ve{s}$. Different from the centralized mechanism, there are no $\ve{\beta}$ among user~3's message components, while user~2 needs to provide two $\ve{\beta}$'s, namely $\ve{\beta}^{2,1},\ve{\beta}^{2,3}$. In addition, for each constraint $l$, every user needs to announce variable~$n$'s to each of her neighbor(s); for each day $t$, every user also needs to provide variable~$\nu$'s to each of her neighbor(s).


For the rest of this subsection, we focus on user~3 and consider how $n$ variables play their roles in the tax evaluation. With this message exchange network, we can write down user~3's tax function explicitly:
\begin{equation*}
       \begin{split}
              \hat{t}^3(m) &= \sum_{t=1}^2 \left(p_t + \Rp^3_t(\ve{s}^{-3},\zeta^{-3})\right) y^3_t - q^{-3,5} y^3_1 - q^{-3,6} y^3_2 + q^{-3,7} (y^3_1 + y^3_2) \\
              &+ \sum_{l=1}^7 \text{pr$\boldsymbol{n}$}^{3,l}(m) + \sum_{t=1}^2 \text{pr$\boldsymbol{\nu}$}^3_t(m) + \sum_{l=1}^7 (q^{3,l} - q^{-3,l})^2 \\
              &+ q^{3,1} (1- n^{2,1,1}) + q^{3,2} (1 - n^{2,1,2}) + q^{3,3} (1 + y^2_1 - n^{2,1,3}) + q^{3,4} (1 + y^2_2 - n^{2,1,4})\\
              &+ q^{3,5} (1 + \beta^2_1 - n^{2,1,5}) + q^{3,6} (1 + \beta^2_2 - n^{2,1,6}) + q^{3,6} (1 + \beta^2_2 - n^{2,1,6}) \\
              &+ q^{3,7} (2 - \beta^2_1 - \beta^2_2 - y^2_1 - y^2_2 - n^{2,1,7}) + \sum_{t=1}^2 \left( (s^3_t - s^{-3}_t)^2 + s^3_t (z^{-i} - \zeta^{-3}_t)\right),
       \end{split}
\end{equation*}
where
\begin{align*}
       \text{pr$\boldsymbol{n}$}^{3,l}(m) &= (n^{3,2,l} - n^{2,1,l})^2, \text{ for } l=1,2,5,6,\\
       \text{pr$\boldsymbol{n}$}^{3,3}(m) &= (n^{3,2,3} + y^2_1 - n^{2,1,3})^2,\\
        \text{pr$\boldsymbol{n}$}^{3,4}(m) &= (n^{3,2,4} + y^2_2 - n^{2,1,4})^2, \\
       \text{pr$\boldsymbol{n}$}^{3,7}(m) &= (n^{3,2,7} - y^2_1 - y^2_2 - n^{2,1,7})^2, \\
       \text{pr$\boldsymbol{\nu}$}^3_t(m) &= (\nu^{3,2}_t - y^2_t - \nu^{2,1}_t)^2,
\end{align*}
and
\begin{align*}
       q^{-3,l} &= q^{2,l},\ l=1,\ldots,7, \\
       s^{-3}_t &= s^2_t,\ t=1,2, \\
       \zeta^{-3}_t &= \beta^{2,3}_t + y^2_t + \nu^{2,1}_t,\ t=1,2,\\
       z^{-3} &= \max \left\{\zeta^{-3}_1, \zeta^{-3}_2 \right\}.
\end{align*}
In user~3's tax function there is no $\text{pr}\boldsymbol{\beta}^3_t(m)$ terms because user~3 is not assigned to any other users for providing $\beta$ proxies.

To figure out how the proxies $n$'s work, here we focus on the 7-th constraint, and see how the corresponding constraint term is evaluated in user 3's tax function. The reason why other $n$'s and $\nu$'s work is similar. For user 3, the constraint term is
\begin{equation*}
       \text{con}^{3,7}(m) = (q^{3,7}-q^{-3,7})^2 + q^{3,7}\underbrace{(2-y^2_1-y^2_2-\beta^{2,3}_1-\beta^{2,3}_2-n^{2,1,7})}_{\text{Slackness part}}.
\end{equation*}
In the centralized mechanism, the slackness part turns out to be $1-\sum_{t=1}^2 \sum_{i=1}^3 x^{i*}_t$ at NE. What we want to show is that with the distributed mechanism, the same outcome can be realized at NE. Same as the centralized mechanism, $\ve{y}^i=\ve{x}^{i*}$ at NE, so $y^2_1=x^{2*}_1,y^2_2=x^{2*}_2$. By Lemma~\ref{lem:dist_eqbeta}, $\beta^{2,3}_1+\beta^{2,3}_2 = y^3_1 + y^3_2 = x^{3*}_1 + x^{3*}_2$, so it remains to show that $n^{2,1,7}=x^{1*}_1+x^{1*}_2$.

Let's trace how the $n^{2,1,7}$ is generated at NE. From \eqref{eqprx:n},
\begin{equation*}
       n^{2,1,7} = a^{1,7}_1 y^1_1 + a^{1,7}_2 y^1_2 + \sum_{h\in \cN(1)\backslash \{2\} } n^{1,h,7}.
\end{equation*}
Notice that $\cN(1)=\{2\}$, so $\cN(1)\backslash\{2\}$ is empty. As a result, at NE, $n^{2,1,7}=y^1_1+y^1_2=x^{1*}_1+x^{1*}_2$.

\subsection{The Learning Algorithm}

Before the algorithm is implemented, one might want to check whether the problem setting satisfies Assumptions~\ref{assump:util} and~\ref{assump:price_set}.

First we can check Assumption~\ref{assump:price_set}. Suppose for the specific environment we have $\underline{r}^i_t=i\cdot t / 9$ and $\bar{r}^i_t = i \cdot t$. Then for the first condition in Assumption~\ref{assump:price_set}, since each $x^i_t$ has a lower bound $-1$, we have
\begin{equation*}
       \dot{v}^i_t(x^i_t) = \frac{i\cdot t}{x^i_t+2}\leq \frac{i\cdot t}{-1+2} = \bar{r}^i_t.
\end{equation*}
Also, every $x^i_t$ is upper bounded by 7 because from the 7-th constraint we have
\begin{equation*}
       x^i_t \leq 2 - \sum_{(i',t')\neq(i,t)} x^{i'}_{t'} \leq 2 - 5 \cdot (-1) = 7,
\end{equation*}
and thus
\begin{equation*}
       \dot{v}^i_t(x^i_t) = \frac{i \cdot t}{x^i_t + 2} \geq \frac{i \cdot t}{7 + 2} = i \cdot t / 9.
\end{equation*}
For the second condition in Assumption~\ref{assump:price_set}, for all $p \in [i\cdot t/9,i\cdot t]$, we have 
\begin{equation*}
      \dot{v}^i_t(x^i_t)=p \Leftrightarrow x^i_t = \frac{i \cdot t}{p} - 2,
\end{equation*}
so Assumption~\ref{assump:price_set} is verified.

With the $\underline{r}^i_t$'s and $\bar{r}^i_t$'s chosen above, a dual feasible set $\cP$ is constructed. Within this price set $\cP$, Algorithm~\ref{algo:dyn} evaluates the function $(\dot{v}^i_t)^{-1}(\cdot)$ only in the interval $[i\cdot t/9,i\cdot t]$. Consequently, in running Algorithm~\ref{algo:dyn} we only need to define $v^i_t(\cdot)$ on the interval $[-1,7]$.

Regarding Assumption~\ref{assump:util}, we need to show that $v^i_t(\cdot)$ is strongly concave on $[-1,7]$. 
Since one can verify that the function $-it \ln(2+x) - ax^2/2$ is convex on $[-1,7]$ for $0\leq a \leq it/81$, every $v^i_t(\cdot)$ is strong concave, and thus Assumption~\ref{assump:util} holds\footnote{This is based on the fact that $f$ is strongly convex with parameter $\delta$ iff $g(\ve{x})=f(\ve{x})-\frac{\delta}{2}\lVert \ve{x} \rVert^2$ is convex.}.

To choose an appropriate step size $\alpha$ for the algorithm, we need to investigate further the parameter $\delta$. In our environment, the sum of utility functions $f(\ve{x})=-\sum_{t=1}^2\sum_{i=1}^3v^i_t(x^i_t)$ is strongly concave on $ [-1,7]^6$ with parameter $\delta=18/81$, because each component $v^i_t$ of $f$ is a strongly concave function with parameter $i\cdot t/81$, and the parameter $\delta$ is additive: $\sum_{t=1}^2 \sum_{i=1}^3 it/81 = 18/81$. By calculation, $\lVert \tilde{A} \rVert \approx 3.1623 $, so one possible step size can be $\alpha=0.1<2\times \delta / \lVert \tilde{A} \rVert$.
According to Algorithm \ref{algo:dyn} the updates required are (define $\eta(i,t)=2(i-1)+t$ for convenience):
\begin{align*}
       &\tilde{q}^{i,\eta(j,t)}(k+1) = q^{i,\eta(j,t)}(k) - \alpha (1+y^j_t(k)),\ j=1,2,3, \ t=1,2, \\
       &\tilde{q}^{i,7} = q^{i,7} - \alpha (2 - \sum_{j=1}^3 \sum_{t=1}^2 y^j_t(k)), \\
       &\tilde{s}^i_t(k+1) = s^i_t(k) + \alpha \sum_{j=1}^3 y^j_t,\ t=1,2, \\
       &(\ve{q}^i(k+1),\ve{s}^i(k+1)) = \mathbf{Proj}_\cP (\tilde{\ve{q}}^i(k+1),\tilde{\ve{s}}^i(k+1)),\\
       &y^i_t(k+1) = \frac{i \cdot t}{p_t - q^{i,\eta(i,t)}(k+1) + q^7(k+1) + s^i_t(k+1)} - 2.
\end{align*}


\begin{figure}
       \centering
       \includegraphics[width=15cm]{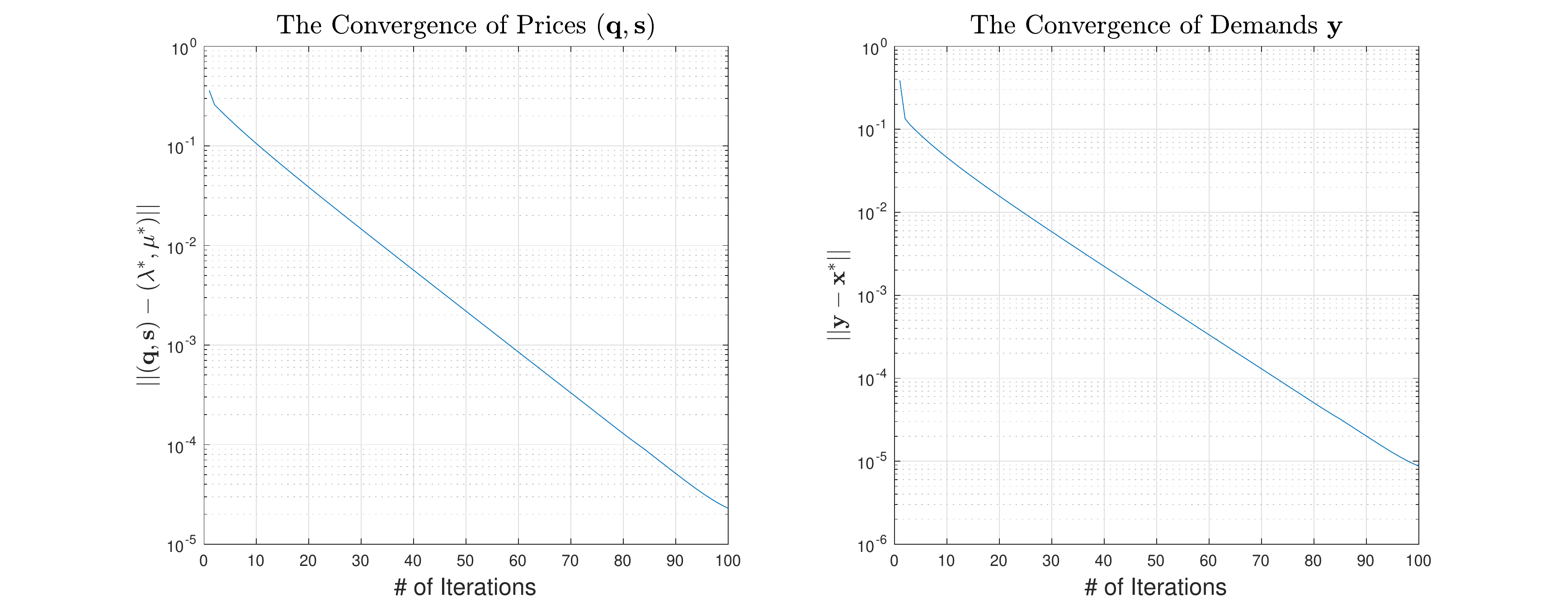}
       \caption{The Convergence of the Learning Algorithm.\label{fig:conv}}
\end{figure}

To verify the convergence of the learning algorithm we run it with initial price  set to $(\ve{q}(0),\ve{s}(0))=\mathbf{Proj}_\cP(\ve{0}_{9\times 1})$. After $K=100$ iterations, we observe the convergence for both the suggested prices $\ve{q},\ve{s}$ and the corresponding announced demands $\ve{y}$. Figure~\ref{fig:conv} shows the process of convergence and verifies that the convergence rate is exponential, as expected.


\section{Conclusions}

Motivated by the work of mechanism design for NUM problems, we proposed a new class of (indirect) mechanisms, 
with application in demand management in energy communities. The proposed mechanisms possess desirable properties including full implementation, individual rationality, and budget balance and can be easily generalized to different environments with peak shaving and convex constraints. 
We showed how the original ``centralized'' mechanism can be modified in a systematic way to account for environments with communication constraints. This modification leads to a new type of mechanisms that we call ``decentralized'' mechanisms and can be thought of as the analog to decentralized optimization (developed for optimization problems with non-strategic agents) for environments with strategic users.
Finally, motivated by the need for practical deployment of these mechanisms, we introduced a PGD-based learning algorithm for users to learn the NE of the mechanism-induced game.

Possible future research directions include  learning algorithms for the distributed mechanism, as well as
co-design of a (distributed) mechanism and characterization of the class of convergent algorithms for this design.

\appendix

\subsection{Equivalence of Centralized Optimization Problem \eqref{opt:woassump} and Original Problem \eqref{opt:central}}
\label{appx:equiv_opt}
We first prove this sufficiency by showing that we can always derive the optimal solution of \eqref{opt:woassump} from the optimal solution of newly constructed \eqref{opt:central}. Suppose the optimal solution of \eqref{opt:central} is $(\ve{x}^*,w^*)$. We claim that $\ve{x}^*$ is the optimal solution of the original problem \eqref{opt:woassump}. First, the feasibility of $\ve{x}^*$ in \eqref{opt:woassump} is assured by \eqref{cp2:feas} in the newly constructed problem.

Now check the optimality. Suppose $\ve{x}^*$ is not the optimal for \eqref{opt:woassump}, and instead, $\ve{x}'$ is the optimal. In the new problem, construct
\begin{equation*}
       \tilde{\ve{x}} = \ve{x}', \quad \tilde{w} = \underset{1 \leq t \leq T}{\max} \sum_{i=1}^N x^{'i}_t,
\end{equation*}
then it is easy to verify that $(\tilde{\ve{x}},\tilde{w})$ is feasible for the new optimization. Notice that
\begin{equation}
       \label{eq:equiv_cp_cp2}
       \begin{split}
              &\sum_{i=1}^N v^i(\tilde{\ve{x}}^i) - \sum_{t=1}^T p_t \sum_{i=1}^N \tilde{x}^i_t - p_0 \tilde{w} \\
              =& \sum_{i=1}^N v^i(\ve{x}^{'i}) - \sum_{t=1}^T p_t \sum_{i=1}^N x^{'i}_t - p_0 \underset{1 \leq t \leq T}{\max} \sum_{i=1}^N x^{'i}_t \\
              >& \sum_{i=1}^N v^i(\ve{x}^{*i}) - \sum_{t=1}^T p_t \sum_{i=1}^N x^{*i}_t - p_0 \underset{1 \leq t \leq T}{\max} \sum_{i=1}^N x^{*i}_t \\
              \geq& \sum_{i=1}^N v^i(\ve{x}^{*i}) - \sum_{t=1}^T p_t \sum_{i=1}^N x^{*i}_t - p_0 w^*.
       \end{split}
\end{equation}
The first inequality follows the optimality of $\ve{x}'$ in the original optimization \eqref{opt:woassump}; the second inequality comes from the constraint \eqref{cp2:peak} in the new optimization. By this inequality chain, we find a $(\tilde{\ve{x}},\tilde{w})$ with a better objective function value in \eqref{opt:central} than $(\ve{x}^*,w^*)$, which contradicts to the assumption that $(\ve{x}^*,w^*)$ is optimal solution of \eqref{opt:central}.

Therefore, by contradiction, we shows that if $(\ve{x}^*,w^*)$ is optimal solution of \eqref{opt:central}, $\ve{x}^*$ must be the optimal solution for the original optimization \eqref{opt:woassump}.

For the other direction, we need to show if $\ve{x}'$ is optimal solution of \eqref{opt:woassump}, then we are able to construct an optimal solution of \eqref{opt:central} based on $\ve{x}'$. We construct $\tilde{\ve{x}} = \ve{x}'$ and $\tilde{w} = \underset{1 \leq t \leq T}{\max} \sum_{i=1}^N x^{'i}_t$ and argue that this $(\tilde{\ve{x}},\tilde{w})$ is the optimal for \eqref{opt:central}. Assume $(\ve{x}^*,w^*)$ is the optimal for \eqref{opt:central}, then we will still get the same inequality chain as \eqref{eq:equiv_cp_cp2} (except that for the second line, there should be a ``greater than or equal'' sign instead), and the equality and inequalities hold for the same reasons as stated above. This shows $(\tilde{\ve{x}},\tilde{w})$ has the same objective value with the optimal solution of \eqref{opt:central}, and therefore $(\tilde{\ve{x}},\tilde{w})$ constructed from $\ve{x}'$ of the original problem is also the optimal for the new problem.

\subsection{Proof of Lemma \ref{lem:kkt1-3}}
\label{appx:pf_lem:kkt1-3}
\begin{IEEEproof}
       At NE $m^*$, for the constraint $l$ in $\cL$, consider the message components $q^{i,l}$ for each user $i$. In user $i$'s tax function, denote the part relative to $q^{i,l}$ by $\hat{t}^{i,l}_q$. We have
       \begin{equation*}
              \begin{split}
                     \hat{t}^{i,l}_q(m^i,m^{-i*}) &= (q^{i,l}-q^{-i,l*})^2 + q^{i,l}\left(b^l - \sum_{j \neq i}\ve{a}^{j,l}\ve{y}^{j*} - \ve{a}^{i,l} \ve{\beta}^{i-1}\right)  \\
                     &=  (q^{i,l}-q^{-i,l*})^2 + q^{i,l} \underbrace{\left(b^l - \sum_{j}\ve{a}^{j,l}\ve{y}^{j*}\right)}_{\text{denoted by } e^l(\ve{y}^*)}\\
                     &(\ve{\beta}^{i-1}=\ve{y}^i \text{ by Lemma \ref{lem:eqbeta}} )
              \end{split}
       \end{equation*}
       For any user $i$, there is no unilateral profitable deviation on $m^{i*}$. Hence, if we fix $m^{-i*}$ and all the message components of $m^{i*}$ except $q^{i,l}$, it is a necessary condition that user $i$ cannot find a better response than $q^{i,l*}$.

       Consider the best response of $q^{i,l}$ in different cases of $e^l(\ve{y}^*)$.

       \textit{Case 1.} $e^l(\ve{y}^*) > 0$, i.e., the constraint $l$ is inactive at NE. Note that $\hat{t}^{i,l}_q$ is a quadratic function of $q^{i,l}$ of the following form
       \begin{equation*}
              \hat{t}^{i,l}_q = (q^{i,l})^2 - (2q^{-i,l*} - e^l(\ve{y}^*))q^{i,l} + (q^{-i,l*})^2.
       \end{equation*}
       Without considering the nonnegative restriction, the best choice should be $q^{-i,l*} - e^l(\ve{y}^*)/2$. Since $q^{i,l} \geq 0$, the best choice for $q^{i,l}$ would be $(q^{-i,l*} - e^l(\ve{y}^*)/2)^+$ (here $(\cdot)^+=\max\{\cdot,0\}$), which is unique with fixed $m^*_{-i}$.

       Therefore,
       \begin{equation*}
              q^{i,l*} = (q^{-i,l*} - e^l(\ve{y}^*)/2)^+.
       \end{equation*}

       Observe that $(q^{-i,l*} - e^l(\ve{y}^*)/2)^+ \leq (q^{-i,l*})^+ = q^{-i,l*}$. Equality holds only if $q^{-i,l*} \leq e^l(\ve{y}^*)/2$ and $q^{-i,l*}=0$. Thus, for all $i$, $q^{i,l*} \leq q^{-i,l*}$, equality holds only if $q^{i,l*}=0$ and $q^{-i,l*}=0$. In other words, if for one user $i$ we have $q^{i,l*} = q^{-i,l*}$, then all the $q^{i,l*}=0$.

       Notice that $q^{i,l*} < q^{-i,l*}$ implies $q^{i,l}$ is smaller than one of the $q^{j,l}$ among user $j\neq i$, which means $q^{i,l}$ is not the largest. Assume that $q^{i,l*} < q^{-i,l*}$ for all $i$, then no $q^{i,l}$ can be the largest among $\{q^{i,l}\}_{i\in \cN}$, but we also know that $\{q^{i,l}\}_{i\in \cN}$ is a finite set and therefore it must have a maximum. Here comes the contradiction. As a result, there must exist at least one $i$, such that $q^{i,l*} = q^{-i,l*}$, which implies that all the $q^{i,l*}=0$.

       \textit{Case 2.} $e^l(\ve{y}^*) = 0$, i.e., the constraint $l$ is active at NE. In this case, $\hat{t}^{i,l}_q = (q^{i,l} - q^{-i,l*})^2$. It is clear that every user's best response is to make her own price align with the average of the others.

       Notice that if $q^{i,l*} = q^{-i,l*}$, then $q^{-i,l}$ is equal to the average of all $q^l$. Consequently, $q^{i,l*} = q^{j,l*}$ for all $i,j \in \cN$.

       \textit{Case 3.} $e^l(\ve{y}^*) < 0$, i.e., the constraint $l$ is violated at NE. In this case,

       \begin{equation*}
              \hat{t}^{i,l}_q = (q^{i,l})^2 - \underbrace{(2q^{-i,l*} - e^l(\ve{y}^*))}_{>0} q^{i,l} + (q^{-i,l*})^2,
       \end{equation*}
       which leads to a condition for all user $i$ as
       \begin{equation*}
              q^{i,l*} = q^{-i,l*} + \underbrace{(-e^l(\ve{y}^*)/2)}_{>0} > q^{-i,l*}.
       \end{equation*}
       In a finite set, if one number is strictly larger than the average of the others, it means it is not the smallest number in the set. If this condition is true for all user $i$, it means there is no smallest number among the set, which is impossible. Therefore, Case 3 won't happen at NE.

       In summary, at NE, we always have $e^l(\ve{y}^*) \geq 0$, and $q^{i,l}$'s are equal. Moreover, $q^{i,l*}e^l(\ve{y}^*)=0$. These prove the primal feasibility, equal prices and complementary slackness on prices $q$ in the Lemma \ref{lem:kkt1-3}.

       Now for the time $t$, consider the message component $s^i_t$ for each user $i$. In user $i$'s tax function, denote the part relative to $s^i_t$ by $\hat{t}^{i,l}_s$. We have
       \begin{equation*}
              \begin{split}
                     \hat{t}^{i,t}_s(m^i,m^{-i*}) &= (s^i_t - s^{-i*}_t)^2 + s^i_t \left(z^{-i}-\sum_{j\neq i}y^{j*}_t - \beta^{i-1*}_t\right) \\
                     &= (s^i_t - s^{-i*}_t)^2 + s^i_t \underbrace{\left(z^* - \sum_j y^{j*}_t\right)}_{\text{denoted by } g_t(\ve{y}^*)},
              \end{split}
       \end{equation*}
       where $z^* = \max_t \sum_j y^{j*}_t$.

       Different from the proof of previous part, here we only need to consider two cases of $g_t(\ve{y}^*)$ because by definition of $z$, $g_t(\ve{y}^*)$ is always nonnegative. Another thing we can observe is that there exists at least one $t$ such that $g_t(\ve{y}^*)=0$, i.e., at least one time $t$ is the time for the peak demand. Define the set $\tilde{\cT}$ as the time set containing all the time $t$'s with peak demand.

       Check the best response of $s_t$ separately. For all the $t' \notin \tilde{\cT}$, $g_{t'}(\ve{y}^*) > 0$, then following the similar steps shown above, we know $s^{i*}_{t'} = 0$ for all $i$. For those $t \in \tilde{\cT}$, we have already had $g_t(\ve{y}^*)=0$. For those $t$, the best response is $s^{i*}_t = s^{-i*}_t$, which is true for every user $i$.
       As a result, we also have the equal prices and complementary slackness for $s^i_t$.

       Check \eqref{comp:wrp}. For $t \in \tilde{\cT}$, $g_t(\ve{y}^*)=0$, so \eqref{comp:wrp} holds.

       For $t \notin \tilde{\cT}$, $g_t(\ve{y}^*)>0$, so
       \begin{equation*}
              t \notin \arg\underset{\tilde{t}}{\max} \sum_{j\neq i} y^j_{\tilde{t}} + \beta^{i-1}_{\tilde{t}}.
       \end{equation*}
       Also, we know such $t \notin \tilde{\cT}$ has $s_t=0$. Therefore, for $t \notin \tilde{\cT}$, for either branch in the definition of radial pricing $\Rp$, $\Rp^i_t(\ve{s},\ve{\zeta}^{-i})=0$. Hence, \eqref{comp:wrp} holds for $t=1,\ldots,T$.
\end{IEEEproof}

\subsection{Proof of Lemma \ref{lem:stat}}
\label{appx:pf_lem:stat}
\begin{IEEEproof}
       At NE, for user $i$, the $u^i(m^i,m^{-i})$ can be treated as a function of $m^i$ with $m^{-i}$ fixed. By the assumption of the existence of NE, $u^i(m^i,m^{-i})$ must have a global maximizer with respect to $m^i$. Given $m^{-i}$, all the auxiliary variables and functions which only determined by $m^{-i}$ are constants here, and one can check that the other terms in (\ref{def:alloc}) and (\ref{def:tax}) are differentiable. Necessary conditions for the global maximizer are
       \begin{equation*}
              \frac{\partial u^i}{\partial y^i_t} = \dot{v}^{i}_t(\hat{x}^i_t(m)) - \left(p_t + \Rp^i_t(\ve{s},\ve{\zeta}^{-i})+\sum_{l\in \cL_i} a^{i,l}_t q^l\right)=0.
       \end{equation*}
       Therefore,
       \begin{equation*}
              \dot{v}^{i}_t(\hat{x}^i_t(m)) = p_t + \Rp^i_t(\ve{s},\ve{\zeta}^{-i})+\sum_{l\in \cL_i} a^{i,l}_t q^l,
       \end{equation*}
       which is exactly the equation \eqref{eq:sta_price}.

       Equation \eqref{eq:sta_peak} follows directly from the definition of $\Rp$ operator.
\end{IEEEproof}

\subsection{Proof of Theorem \ref{thm:exist}}
\label{appx:pf_thm:exist}
\begin{IEEEproof}
       By assumption, the centralized problem is a convex optimization problem with non-empty feasible set, so there must exist an optimal solution $\left\{\ve{x}^*,w^*\right\}$ and corresponding Lagrange multipliers $\lambda^{l*}, \mu_t^*$ which satisfy KKT conditions (\ref{KKT:pfeas_x}-\ref{KKT:stat_cons}).

       Consider the message profile $m^*$ consisting of
       \begin{align*}
              y^i_t &= x^{i*}_t,\ t=1,\ldots,T,\ \forall i \in \cN, \\
              q^{i,l} &= \lambda^{l*},\ l \in \cL_i,\ \forall i \in \cN, \\
              s^i_t &= \mu_t^*,\ t=1,\ldots,T,\ \forall i \in \cN\\
              \beta^i_t &= x^{i+1*}_t,\ t=1,\ldots,T,\ \forall i \in \cN.
       \end{align*}
       If for arbitrary user $i$, no profitable unilateral deviations exist, i.e., there does not exist an $\tilde{m}=\left(\tilde{m}_i,m_{-i}^*\right)$, such that $u_i(\tilde{m}) > u_i(m^*)$, then $m^*$ is a NE of the game $\cG$.

       We can focus on $u_i(m)$ of user $i$, to see whether she has a profitable deviation given $m_{-i}^*$. For user $i$, we have
       \begin{align*}
              &q^{-i,l} = \lambda^{l*},\ \forall l \in \cL\\
              &s^{-i}_t = \mu^*_t,\ t=1,\ldots,T, \\
              &\Rp^i_t(\ve{s}^{-i},\ve{y}^{-i},\ve{\beta}^{i-1}) = \mu^*_t,\ t=1,\ldots,T,\\
              &z^{-i} = w^* = \underset{t}{\max} \left(\sum\nolimits_{j\in \cN} x^{j*}_t\right).
       \end{align*}

       Therefore, in the interest of user $i$, she wants to maximize the following
       \begin{align}
              \label{eq:proof_exist_sim}
              \begin{split}
                     &u^i(m^i,m^{-i*}) =\\
                     & \sum_{t=1}^T \underbrace{ \left(v^i_t(y^i_t) - (p_t+\mu_t^*)y^i_t - \sum_{l \in \cL_i} \lambda^{l*} a^{i,l}_t y^i_t\right)}_{\text{Function of } y^i_t} \\
                     &- \sum_{l \in \cL} \underbrace{\left((q^{i,l}-\lambda^{l*})^2+q^{i,l}\left(b^l-\sum_j \sum_{t=1}^T a^{j,l}_t x^{j*}_t\right)\right)}_{\text{Function of } q^{i,l}}\\
                     & - \sum_{t=1}^T \underbrace{\left((s^i_t-\mu_t^*)^2+s^i_t \left(w^* - \sum_j x^{j*}_t\right)\right) }_{\text{Function of } s^i_t} \\
                     & - \sum_{t=1}^T \underbrace{(\beta^i_t-x^{i+1*}_t)^2}_{\text{Function of } \beta^i_t}.
              \end{split}
       \end{align}
       The last term of (\ref{eq:proof_exist_sim}) is the only term related to $\ve{\beta}^i$, which is a quadratic terms. As a strategic agent, it is clear that user $i$ won't deviate from $\ve{\beta}^i=\ve{x}^{i+1*}$ otherwise she will pay for the penalty from this.

       The second and third terms of (\ref{eq:proof_exist_sim}) are quite similar: they both consist of a quadratic term and a term for complementary slackness. For the second term, let's consider constraint $l$. If $l$ is active in the optimal solution, the complementary slackness term goes to 0. To avoid extra payment, user $i$ will not deviate $q^{i,l}$ from the price suggested by optimal solution $\lambda^{l*}$. If $l$ is inactive, the price $\lambda^{l*}$ suggested by optimal solution will be 0. Then the penalty of constraint $l$ for user $i$ is
       \begin{equation*}
              (q^{i,l})^2 + q^{i,l} \underbrace{\left(b^l - \sum_j \sum_{t=1}^T a^{j,l}_t x^{j*}_t\right)}_{>0},
       \end{equation*}
       where user $i$ can only select a nonnegative price $q^{i,l}$. There are no better choices better than choosing $q^{i,l}=0=\lambda^{l*}$. Similar analysis works for the third term of (\ref{eq:proof_exist_sim}). As a result, there are not unilateral profitable deviations on $q^{i,l}$ for all $l$, and $s^i_t$ for all $t$.

       Now we denote the terms in the parentheses in the first part of (\ref{eq:proof_exist_sim}) by $f^i_t(y^i_t)$. Since these four terms are disjoint in the aspect of inputted variables, $u^i(m^i,m^{-i*})$ achieves its maximum if and only if every $f^i_t(y^i_t)$ achieves its maximum, and the rest three terms equal their minimum. As for the first part, due to the strict concavity of $v^i_t(\cdot)$, the second order derivative of $f^i_t(y^i_t)$ for each $t$ is negative, which indicates that $f^i_t(y^i_t)$ is strictly concave as well. We can find the maxima of $f^i_t(y^i_t)$ by first order condition:
       \begin{equation}
              \label{eq:proof_exist_f}
              \frac{\mathrm{d}f^i_t}{\mathrm{d}y^i_t}(y^{i*}_t)=\dot{v}^i_t(y^i_t) - (p_t+\mu_t^*) - \sum_{l \in \cL_i} \lambda^{l*} a^{i,l}_t = 0.
       \end{equation}
       By \eqref{KKT:stat_cons} in the KKT conditions, we know the only $y^i_t$ that makes \eqref{eq:proof_exist_f} hold is $y^i_t= x^{i*}_t$ for all $t$. The reason is that by the strict concavity assumption of utility function $v^i_t(\cdot)$, the first order derivative of $v^i_t$ is strictly decreasing and therefore, for one aggregated price, there is at most one demand value $x$ that makes $\dot{v}^i_t(x)$ equals that price.

       Therefore, for any agent $i$, if others send messages $m^{-i*}$, the only best response of agent $i$ is to announce $m^{i*}$. Under this circumstance, sending messages other than $m^{i*}$ won't increase agent $i$'s payoff $u^i(m)$. Consequently, $m^*$ is a NE of the induced game $\cG$.
\end{IEEEproof}

\subsection{Proof of Theorem \ref{thm:ir}}
\label{appx:pf_thm:ir}
\begin{IEEEproof}
       For any user $i$, if she chooses to participate with other users, when every one anticipates the NE, user $i$'s payoff is of the form \eqref{eq:final_payoff} if she only considers to modify $\ve{y}^i$ and keeps other components unchanged. Thus, user $i$ is facing the following optimization problem
       \begin{equation*}
            \ve{y}^i = \arg \underset{\ve{y}^i \in \Real^T}{\max} \quad v^i(\ve{y}^i) - \sum_{t=1}^T \left(p_t + \Rp^i_t(\ve{s},\ve{\zeta}^{-i})\right) y^i_t - \sum_{l\in \cL_i} q^{-i,l}\sum_{t=1}^T a^{i,l}_t y^i_t.
       \end{equation*}

       By the definition of NE, $\ve{y}^{i*}$ is one of the best solutions, which yields a payoff $u^i(m^*)$. User $i$ can also choose $\tilde{\ve{y}}^i=\ve{0}$. Denote the corresponding message by $\tilde{m}_i$. Then, the payoff value becomes $u^i(\tilde{m}^i,m^{-i*})=v^i(\ve{0})$, which coincides with the payoff for not to participate. Since $m^{i*}$ is the best response to $m^{-i*}$, we have $u^i(m^*) \geq u^i(\tilde{m}^i,m^{-i*}) = v^i(\ve{0})$. In other words, if every one anticipates the NE as the outcome, to participate is at least no worse than not to participate.
\end{IEEEproof}

\subsection{Proof of Theorem \ref{thm:bb}}
\label{appx:pf_thm:bb}
\begin{IEEEproof}
       Suppose the optimal solution for the original problem given by NE is $(\ve{x}^*,\ve{\lambda}^*,\ve{\mu}^*)$, then the tax for user $i$ is
       \begin{equation*}
            \hat{t}^i (m^*) - J(\hat{\ve{x}}^i_t(m^*)) = \sum_{t=1}^T (p_t + \mu^*_t) x^{i*}_t + \sum_{l\in \cL_i} \lambda^{l*}\sum_{t=1}^T a^{i,l}_t x^{i*}_t - J(\ve{x}^{i*}).
       \end{equation*}
       The total amount of tax is
       \begin{equation*}
              \begin{split}
                     &\sum_{i\in \cN} \hat{t}^i (m^*) - J(\ve{x}^{i*}) \\
                     =& \sum_{i \in \cN} \sum_{t=1}^T (p_t + \mu^*_t) x^{i*}_t + \sum_{i \in \cN} \sum_{l\in \cL_i} \lambda^{l*}\sum_{t=1}^T a^{i,l}_t x^{i*}_t - J(\ve{x}^{i*})\\
                     =& \sum_{t=1}^T \left(p_t \sum_{i \in \cN}x^{i*}_t+ \mu^*_t \sum_{i \in \cN}x^{i*}_t \right) + \sum_{l\in \cL} \lambda^{l*}\sum_{t=1}^T \sum_{i \in \cN} a^{i,l}_t x^{i*}_t - J(\ve{x}^{i*}) \\
                     =& \sum_{l\in \cL} \lambda^{l*}\sum_{t=1}^T \sum_{i \in \cN} a^{i,l}_t x^{i*}_t.
              \end{split}
       \end{equation*}

       For each constraint $l$, by the complementary slackness, we have
       \begin{equation*}
              \lambda^{l*} \left(b^l - \sum_{t=1}^T \sum_{i \in \cN} a^{i,l}_t x^{i*}_t\right) = 0.
       \end{equation*}
       Therefore,
       \begin{equation*}
              \sum_{i\in \cN} \hat{t}^i (m^*) - J(\ve{x}^{i*}) = \sum_{l \in \cL} \lambda^{l*} b^l \geq 0,
       \end{equation*}
       which shows that at NE, the planner's payoff is nonnegative.

       Furthermore, in order to save unnecessary expenses on the planner, the energy community can adopt the mechanism with the following tax function $\tilde{t}^i(m)$ instead
       \begin{equation*}
              \tilde{t}^i(m) = \hat{t}^i(m) - \sum_{l\in \cL} q^{-i,l} b^l / N.
       \end{equation*}
       Note that user $i$ has no control on the additional term because no components of $m^i$ are in that term, and thus the additional term won't change NE. Since the prices are equal at NE, so the planner actually gives $\sum_{l \in \cL}\lambda^{l*}b^l$ back to the users. Hence,
       \begin{equation*}
              \sum_{i\in \cN} \tilde{t}^i (m^*) - J(\ve{x}^{i*}) = 0,
       \end{equation*}
       As a side comment, the choice of $\tilde{t}^i(m)$ is not unique. Any adjustment works here as long as it does not depend on $m^i$ for each $t^i(\cdot)$, and sums up to $\sum_{l \in \cL}\ \lambda^{l*}b^l$ at NE.
\end{IEEEproof}

\subsection{Proof of Lemma \ref{lem:mes_comm}}
\label{appx:pf_lem:mes_comm}
\begin{IEEEproof}
       Here we provide a non-rigorous proof of (\ref{eqprx:nu:ctr}). The proof of (\ref{eqprx:n:ctr}) is quite similar. For the detailed version of the proof, we refer the interested readers to 7.1, Chapter 4 of \cite{sinha2017mechanism2}.

       Before we show the proof of this part, for the sake of convenience, we define $n(i,k)$ as the nearest user among the neighbors of user $i$ and user $i$ itself to user $k$. $n(i,k)$ is well-defined because one can show that $n(i,k)=j$ provides a partition for all the users.

       (\ref{eqprx:nu:ctr}) can be shown by applying (\ref{eqprx:nu}) iteratively. Recall that the message exchange network is assumed to be a undirected acyclic graph (i.e. a tree). First consider the user $j$ on the leaves (the nodes with only one degree). Suppose the neighbor of user $j$ is $i$, then $\cN(j)=\{i\}$. By (\ref{eqprx:nu}), we have $\nu^{i,j}_t = y^j_t$. Since no $k$ satisfies $n(i,k)=j$ other than $j$ herself, (\ref{eqprx:nu:ctr}) holds for $\nu^{i,j}_t$ where $j$ is a leaf node.

       For more general cases, to compute $\nu^{i,j}_t$, it is safe to only consider the subgraph $\Gr_i$ contains only node $i$ and node $k$'s such that $n(i,k)=j$. When applying (\ref{eqprx:nu}), it is impossible to have node $l \in \Gr_i^C$ involve, because if it happens when expanding ``$\nu$'' term for some $j'$, $l$ is a neighbor of $j'$. We know that there is a route from $i$ to $j'$, say, route $iLj'$. Since $l\in \Gr_i^C$, $n(i,l) \neq j$, there exists a route $L'$ does not involve any node in branch starting from node $j$, such that $lL'i$, which results in a loop $lL'iLj'l$.

       Then by using (\ref{eqprx:nu}) iteratively, we can see that: 1. every node in $\Gr_i$ will be visited at least once and gives a corresponding demand ``$y$''; 2. each $y^j_t$ is given only once (except the root $i$, who won't give $y^i_t$ in this procedure); 3. when it proceeds to the leaf nodes, the iteration terminates because there are no more ``$\nu$'' terms to expand. Hence, $\nu^{i,j}_t=\sum_{h\in\Gr_i}y^h_t$, and we can easily verify that $\Gr_i\backslash\{i\}$ is nothing but $\{h:n(i,h)=j\}$.
\end{IEEEproof}

\subsection{Convergence of the Learning Algorithm for Centralized Mechanism}
\label{appx:pf_conv}
The convergence of the proposed learning algorithm can be shown in three steps mentioned in Section \ref{sec:learn}. First step shows the connection between $m^*$ and $x^*,\ve{\lambda}^*,\ve{\mu}^*$ of the optimal solution for the original optimization, which has already been clarified in Section \ref{sec:learn}. As a result, learning NE is equivalent to learning the optimal solution of the original optimization problem. For the second step, as a convex optimization problem with non-empty feasible set defined by linear inequalities, Slater's condition is easy to check. Therefore, we have strong duality in this problem, which means we can obtain the optimal solution of the original problem as long as we solve the dual problem. The last step is to identify the dual problem and find a convergent algorithm for it. This part of appendix explains how to pin down the dual function and the dual feasible set, and shows the convergence of PGD algorithm on this dual problem.

Before we identify the dual function of the original problem, for the sake of convenience, in constraint \eqref{cp2:peak} of the original problem, move $w$ to the left hand side, and rewrite \eqref{cp2:feas}
\eqref{cp2:peak} into one matrix form
\begin{equation*}
       \tilde{A}\ve{x} + \tilde{\ve{1}} w \leq \tilde{\ve{b}},
\end{equation*}
where $\tilde{A}$ is defined in \eqref{def:tilde}, and
\begin{equation*}
       \tilde{\ve{1}} = \left(
       \begin{array}{c}
              \ve{0}_L \\
              -\ve{1}_T
       \end{array}
       \right), \
       \tilde{\ve{b}} = \left(
              \begin{array}{c}
                     \ve{b} \\
                     \ve{0}_{T}
              \end{array}
       \right).
\end{equation*}
 Suppose $f(\ve{x})=\sum_i \sum_t (v^i_t(x^i_t)-p_t x^i_t)$, then the objective function can be written as $f(\ve{x})-p_0 w$. Observe that by Assumption \ref{assump:util}, $(v^i_t(x^i_t)-p_t x^i_t)$'s are also strongly concave without cross terms. Sequently, one can show directly by the definition of strong concavity that as the sum of these strongly concave functions, $f(\ve{x})$ is strongly concave as well. Let $h(\ve{x})=-f(\ve{x})$, then $h(\ve{x})$ is strongly convex with parameter $\delta'$. Denote by $h^*(\cdot)$ the conjugate function of $h(\ve{x})$.

 With these notations in mind, the dual function of the original problem is
 \begin{equation*}
       \begin{split}
              D(\tilde{\ve{\lambda}}) =& \sup_{\ve{x},w} \left\{f(\ve{x}) - p_0 w -  \tilde{\ve{\lambda}}^\Tr (\tilde{A}\ve{x} + \tilde{\ve{1}} w - \tilde{\ve{b}})\right\} \\
              =& \ve{b}^\Tr \ve{\lambda} + \sup_{\ve{x}} \left\{(-\tilde{A}^\Tr \tilde{\ve{\lambda}})^\Tr \ve{x} - h(\ve{x})\right\} + \sup_w \left\{\ve{1}_T^\Tr \ve{\mu} w - p_0 w\right\}\\
              =& \ve{b}^\Tr \ve{\lambda} + h^*(-\tilde{A}^\Tr \tilde{\ve{\lambda}}),
       \end{split}
\end{equation*}
Here we should be cautious about the domain of $D(\tilde{\ve{\lambda}})$. In the second line, $\sup_w \{\ve{1}_T^\Tr \ve{\mu} w - p_0 w\}$ is only defined when the coefficient $\ve{1}_T^\Tr \ve{\mu} - p_0=0$, i.e., $\sum_t \mu_t = p_0$. Therefore, we get the following dual problem.
\begin{align}
       \underset{\ve{\lambda},\ve{\mu}}{\text{minimize}} \quad& \ve{b}^\Tr \ve{\lambda} + h^*(-\tilde{A}^\Tr \tilde{\ve{\lambda}}) \\
       \label{dual:con_p}
       \text{subject to} \quad & \sum_t \mu_t = p_0, \\
       \label{dual:con_nonneg}
       &\ve{\lambda} \geq 0, \ \ve{\mu} \geq 0.
\end{align}

Now we have derived the dual problem for the original optimization. To find the optimal solution, one direct thought is to use projected gradient descent.
Luckily, we have the following theorem which ensures the convergence of PGD algorithm.
\begin{theorem}
       \label{thm:PGDconv}
       For a minimization problem on a closed and convex feasible set $\cX$ with objective function $f(\ve{x})$, suppose $\cX^*$ is the set of optimal solutions. If $f$ is convex and $\beta$-smooth on $\cX$, by using PGD with step size $\alpha < 2/\beta$, there exists $\ve{x}^* \in \cX^*$, such that
       \begin{equation*}
              \lim_{k\to \infty} \ve{x}(k) = \ve{x}^*.
       \end{equation*}
\end{theorem}
\begin{IEEEproof}
       The proof can be found in \cite[Thm.~1, Sec.~7.2]{polyak1987opt}.
\end{IEEEproof}

Theorem \ref{thm:PGDconv} indicates that if the dual problem satisfies certain conditions, the solution converges to the set of optimal solutions. Although it is not clear whether the dual problem has a unique solution, by strong duality and the uniqueness of the solution to the primal problem, no matter which dual optimal solution is achieved, the corresponding primal solution can only be the unique optimal one, and results in the same outcome.

Now for the dual problem, we need to check the conditions required by Theorem \ref{thm:PGDconv}. First check the objective function. It is clear that any conjugate functions are convex, so $h^*(\cdot)$ is convex, and consequently $h^*(-\tilde{A}\tilde{\ve{\lambda}})$ is convex in $\tilde{\ve{\lambda}}$ as a composition of convex function and affine function. Thus, the objective function is convex. Since $h(\ve{x})$ is strongly convex with parameter $\delta'$, by the result mentioned in \cite{zhou2018fenchel}, $\delta'$-strong convexity of $h(\cdot)$ implies $1/\delta'$-smooth of its conjugate $h^*(\cdot)$. Then we have
\begin{equation*}
       ||\nabla h^*(-\tilde{A}^\Tr \tilde{\ve{\lambda}}^1) - \nabla h^*(-\tilde{A}^\Tr \tilde{\ve{\lambda}}^2)|| \leq 1/\delta' \cdot ||-\tilde{A}^T(\tilde{\ve{\lambda}}^1-\tilde{\ve{\lambda}}^2)|| \leq (\|\tilde{A}\|/\delta') \cdot ||\tilde{\ve{\lambda}}^1-\tilde{\ve{\lambda}}^2||,
\end{equation*}
which indicates that the objective function is $\beta$-smooth with $\beta = \|\tilde{A}\|/\delta'$. However, we are not sure whether the objective function is well-defined on the whole feasible set. Fortunately, by Assumption \ref{assump:price_set}, we know that the optimal price vector $\tilde{\lambda}^*$ lies in $\cP$, and we can verify that $\cP$ is a subset of the feasible set generated by \eqref{dual:con_p}\eqref{dual:con_nonneg}. Therefore, by solving the following optimization problem we can obtain the same optimal solution, and Theorem \ref{thm:PGDconv} is applicable here.
\begin{equation}
       \label{dual:withp}
       \underset{\tilde{\ve{\lambda}}\in \cP}{\text{minimize}} \quad \ve{b}^\Tr \ve{\lambda} + h^*(-\tilde{A}^\Tr \tilde{\ve{\lambda}}).
\end{equation}

Apply PGD to \eqref{dual:withp} with step size $\alpha \leq 2/\beta = 2\delta' / \|\tilde{A}\|$, the update rules are as follows:
\begin{align}
       \label{PGD:lambda}
       &\hat{\lambda}^l(k+1) = \lambda^l(k) - \alpha \left( b^l + [- \tilde{A} \nabla h^*(-\tilde{A}^\Tr \tilde{\ve{\lambda}}(k))]_l\right), \\
       \label{PGD:mu}
       &\hat{\mu}_t(k+1) = \mu(k) - \alpha [- \tilde{A} \nabla h^*(-\tilde{A}^\Tr \tilde{\ve{\lambda}}(k))]_{L+t}, \\
       \label{PGD:proj}
       &(\ve{\lambda}(k+1),\ve{\mu}(k+1)) = \mathbf{Proj}_\cP \left(\hat{\lambda}(k+1),\hat{\ve{\mu}}(k+1)\right) .
\end{align}
where $[\cdot]_j$ represents the $j$-th entry of inputted vector. To modify these rules into a learning algorithm for centralized mechanism, by the relation between $m^*$ and the optimal solution of the original problem and the dual, one might want to substitute $\ve{\lambda}, \ve{\mu}$ with $\ve{q}^i, \ve{s}^i$ for each user $i$. However, $\nabla h^*$ is not tractable for users as they do not know the utilities of the others. Thankfully, users can obtain the values of $\nabla h^*$-related terms by cooperation without revealing their entire utility functions. This way is realized by inquiries for the demands under given prices from each user. A key point for this implementation is to build a connection between $\nabla h^*$ and the marginal value function $\dot{v}^i_t$ for each demand $x^i_t$.

A useful result of subgradient of function $f$ and its conjugate $f^*$ can be used here, which is quoted as Theorem \ref{thm:subgrad}.
\begin{theorem}
       \label{thm:subgrad}
      Suppose $f^*(\ve{s})$ is the conjugate of $f(\ve{x})$, then
      \begin{equation*}
             \ve{x} \in \partial f^*(\ve{s}) \Leftrightarrow \ve{s} \in \partial f(\ve{x}).
      \end{equation*}
\end{theorem}
\begin{IEEEproof}
       The proof can be found in \cite{zhou2018fenchel}.
\end{IEEEproof}
Since $h$ is closed (because $h$ is proper convex and continuous) and strictly convex by assumption, $h^*$ is differentiable (see \cite{zhou2018fenchel}) and therefore the subgradient of $h^*$ on a fixed $\ve{p}$ is a singleton. As a result,
\begin{equation*}
       \ve{x}=\nabla h^*(-\ve{\rho}) \Leftrightarrow \ve{\rho}=-\nabla h(\ve{x}) = \nabla f(\ve{x})\Leftrightarrow x^i_t = (\dot{v}^{i}_t)^{-1} (p_t + \rho^i_t).
\end{equation*}
The last equivalent sign comes from the fact that
\begin{equation*}
       [\nabla f(\ve{x})]_{(i-1)T+t} =  \frac{\mathrm{d}}{\mathrm{d}x^i_t}(v^i_t(x^i_t)-p_t x^i_t) = \dot{v}^i_t(x^i_t) - p_t.
\end{equation*}
Thus, in every iteration of PGD, before doing \eqref{PGD:lambda}\eqref{PGD:mu}, one can first evaluate
\begin{equation}
       \label{PGD:x}
       x^i_t(k) = (\dot{v}^i_t)^{-1}(p_t + [\tilde{A}^\Tr \tilde{\ve{\lambda}(k)}]_{(i-1)T+t}),
\end{equation}
and then \eqref{PGD:lambda}\eqref{PGD:mu} become
\begin{align}
       \label{PGD:new_lambda}
       &\hat{\lambda}^l(k+1) = \lambda^l(k) - \alpha \left( b^l - [ \tilde{A} \ve{x}(k)]_l\right) = \lambda^l(k) - \alpha \left( b^l - \ve{a}^l \ve{x}(k)\right), \\
       \label{PGD:new_mu}
       &\hat{\mu}_t(k+1) = \mu(k) + \alpha [ \tilde{A} \ve{x}(k)]_{L+t} = \mu(k) + \alpha \sum_j x^j_t(k).
\end{align}

Arranging \eqref{PGD:proj}, \eqref{PGD:x}, \eqref{PGD:new_lambda} and \eqref{PGD:new_mu} in an appropriate order, we get an algorithm with the same convergent property of the original PGD, and significantly, no $\nabla h^*$ is in the algorithm. By substituting $\ve{\lambda}$ and $\ve{\mu}$ with $\ve{q}^i$ and $\ve{s}^i$ (by making duplications of \eqref{PGD:new_lambda}, \eqref{PGD:new_mu} for each user $i$), and substituting $\ve{x}$ with $\ve{y}$, we obtain Algorithm \ref{algo:dyn}. Consequently, Theorem \ref{thm:conv} follows directly from the convergence of PGD indicated in Theorem \ref{thm:PGDconv}.

\bibliographystyle{IEEEtran}
\bibliography{IEEEabrv,achilleas18abrv,achilleas19_own,abhinav,xupeng}
\end{document}